\documentclass[12pt]{article}
\usepackage{epsfig}
\usepackage{amsmath}
\usepackage{hhline}
\usepackage{amssymb}
\usepackage{times}

\newlength{\dinwidth}
\newlength{\dinmargin}
\begin{document}  
% The rest
\newcommand{\pom}{{I\!\!P}}
\newcommand{\slowpi}{\pi_{\mathit{slow}}}
\newcommand{\fiidiii}{F_2^{D(3)}}
\newcommand{\fiidiiiarg}{\fiidiii\,(\beta,\,Q^2,\,x)}
\newcommand{\n}{1.19\pm 0.06 (stat.) \pm0.07 (syst.)}
\newcommand{\nz}{1.30\pm 0.08 (stat.)^{+0.08}_{-0.14} (syst.)}
\newcommand{\fiidiiiful}{F_2^{D(4)}\,(\beta,\,Q^2,\,x,\,t)}
\newcommand{\fiipom}{\tilde F_2^D}
\newcommand{\ALPHA}{1.10\pm0.03 (stat.) \pm0.04 (syst.)}
\newcommand{\ALPHAZ}{1.15\pm0.04 (stat.)^{+0.04}_{-0.07} (syst.)}
\newcommand{\fiipomarg}{\fiipom\,(\beta,\,Q^2)}
\newcommand{\pomflux}{f_{\pom / p}}
\newcommand{\nxpom}{1.19\pm 0.06 (stat.) \pm0.07 (syst.)}
\newcommand {\gapprox}
   {\raisebox{-0.7ex}{$\stackrel {\textstyle>}{\sim}$}}
\newcommand {\lapprox}
   {\raisebox{-0.7ex}{$\stackrel {\textstyle<}{\sim}$}}
\def\gsim{\,\lower.25ex\hbox{$\scriptstyle\sim$}\kern-1.30ex%
\raise 0.55ex\hbox{$\scriptstyle >$}\,}
\def\lsim{\,\lower.25ex\hbox{$\scriptstyle\sim$}\kern-1.30ex%
\raise 0.55ex\hbox{$\scriptstyle <$}\,}
\newcommand{\pomfluxarg}{f_{\pom / p}\,(x_\pom)}
\newcommand{\dsf}{\mbox{$F_2^{D(3)}$}}
\newcommand{\dsfva}{\mbox{$F_2^{D(3)}(\beta,Q^2,x_{I\!\!P})$}}
\newcommand{\dsfvb}{\mbox{$F_2^{D(3)}(\beta,Q^2,x)$}}
\newcommand{\dsfpom}{$F_2^{I\!\!P}$}
\newcommand{\gap}{\stackrel{>}{\sim}}
\newcommand{\lap}{\stackrel{<}{\sim}}
\newcommand{\fem}{$F_2^{em}$}
\newcommand{\tsnmp}{$\tilde{\sigma}_{NC}(e^{\mp})$}
\newcommand{\tsnm}{$\tilde{\sigma}_{NC}(e^-)$}
\newcommand{\tsnp}{$\tilde{\sigma}_{NC}(e^+)$}
\newcommand{\st}{$\star$}
\newcommand{\sst}{$\star \star$}
\newcommand{\ssst}{$\star \star \star$}
\newcommand{\sssst}{$\star \star \star \star$}
\newcommand{\tw}{\theta_W}
\newcommand{\sw}{\sin{\theta_W}}
\newcommand{\cw}{\cos{\theta_W}}
\newcommand{\sww}{\sin^2{\theta_W}}
\newcommand{\cww}{\cos^2{\theta_W}}
\newcommand{\trm}{m_{\perp}}
\newcommand{\trp}{p_{\perp}}
\newcommand{\trmm}{m_{\perp}^2}
\newcommand{\trpp}{p_{\perp}^2}
\newcommand{\alp}{\alpha_s}

\newcommand{\alps}{\alpha_s}
\newcommand{\sqrts}{$\sqrt{s}$}
\newcommand{\LO}{$O(\alpha_s^0)$}
\newcommand{\Oa}{$O(\alpha_s)$}
\newcommand{\Oaa}{$O(\alpha_s^2)$}
\newcommand{\PT}{p_{\perp}}
\newcommand{\JPSI}{J/\psi}
\newcommand{\sh}{\hat{s}}
\newcommand{\uh}{\hat{u}}
\newcommand{\MP}{m_{J/\psi}}
\newcommand{\PO}{I\!\!P}
\newcommand{\xbj}{x}
\newcommand{\xpom}{x_{\PO}}
\newcommand{\ttbs}{\char'134}
\newcommand{\xpomlo}{3\times10^{-4}}  
\newcommand{\xpomup}{0.05}  
\newcommand{\dgr}{^\circ}
\newcommand{\pbarnt}{\,\mbox{{\rm pb$^{-1}$}}}
\newcommand{\gev}{\,\mbox{GeV}}
\newcommand{\WBoson}{\mbox{$W$}}
\newcommand{\fbarn}{\,\mbox{{\rm fb}}}
\newcommand{\fbarnt}{\,\mbox{{\rm fb$^{-1}$}}}
%
% Some useful tex commands
%
\newcommand{\qsq}{\ensuremath{Q^2} }
\newcommand{\gevsq}{\ensuremath{\mathrm{GeV}^2} }
\newcommand{\et}{\ensuremath{E_t^*} }
\newcommand{\rap}{\ensuremath{\hat{y}} }
\newcommand{\gp}{\ensuremath{\gamma^*}p }
\newcommand{\dsiget}{\ensuremath{{\rm d}\sigma_{ep}/{\rm d}E_t^*} }
\newcommand{\dsigrap}{\ensuremath{{\rm d}\sigma_{ep}/{\rm d}\eta^*} }
% Journal macro
\def\Journal#1#2#3#4{{#1} {\bf #2}, #3 (#4)}
\def\NCA{\em Nuovo Cimento}
\def\NIM{\em Nucl. Instrum. Methods}
\def\NIMA{{\em Nucl. Instrum. Methods} A}
\def\NPB{{\em Nucl. Phys.} B}
\def\PLB{{\em Phys. Lett.}  B}
\def\PRL{\em Phys. Rev. Lett.}
\def\PRD{{\em Phys. Rev.} D}
\def\ZPC{{\em Z. Phys.} C}
% here is an example how to use it :
% \bibitem{ellis}R.K. Ellis, and P. Nason, \Journal{\NPB}{312}{551}{1989}.

\newcommand{\ra}{\rightarrow}
\newcommand{\ccb}{c\bar{c}}
\newcommand{\bbb}{b\bar{b}}
\newcommand{\ftc}{F_2^{c}}

%===============================title page=============================
\begin{titlepage}

\noindent 
DESY 98--204    \hfill ISSN 0418--9833
\newline\noindent
December 1998

\vspace{2cm}
\begin{center}
\begin{Large}
{\bf  Measurement of \boldmath $D^{\ast}$ \unboldmath Meson Cross Sections}
{\bf  at HERA and} \\
{\bf  Determination of the Gluon Density in the Proton} \\ 
{\bf  using NLO QCD} \\

\vspace{2cm}

H1 Collaboration

\end{Large}
\end{center}

\vspace*{4cm}

\begin{abstract}\noindent
With the H1 detector at the $ep$ collider HERA, 
$D^{\ast}$ meson production cross sections
have been measured 
in deep inelastic scattering 
with four-momentum transfers $Q^2>2$ GeV$^2$
and in photoproduction at energies around 
$W_{\gamma p}\approx 88$ GeV and $194$ GeV.
Next-to-Leading Order QCD calculations are found to describe
the differential cross sections within theoretical 
and experimental uncertainties. 
Using these calculations,
the NLO gluon momentum distribution in the proton, 
$x_g g(x_g)$, has been extracted 
in the momentum fraction range $7.5\cdot 10^{-4}< x_g <4\cdot 10^{-2}$
at average scales $\mu^2 =$ 25 to 50 GeV$^2$.
The gluon momentum fraction $x_g$ 
has been obtained
from the measured kinematics of the scattered electron and the 
$D^{\ast}$ meson in the final state.
The results compare well with the gluon distribution obtained from 
the analysis of scaling violations of the proton structure function $F_2$.
\end{abstract}

\vspace{1.5cm}
 
\centerline{\it Submitted to Nuclear Physics B }

\end{titlepage}
\begin{flushleft}
  %   H1AUTS  Author list by names, no. of authors  357
%           status: 02/09/98   08.32.17
 C.~Adloff$^{34}$,                %WUPP-ST                  Adloff              
 M.~Anderson$^{22}$,              %MANC-LEFT  10/97         Anderson            
 V.~Andreev$^{25}$,               %LPI -PD                  Andreev             
 B.~Andrieu$^{28}$,               %ECPL-PD                  Andrieu             
 V.~Arkadov$^{35}$,               %ZEUT-ST    10/96         Arkadov             
 C.~Arndt$^{11}$,                 %DESY-LEFT  10/97         Arndt               
 I.~Ayyaz$^{29}$,                 %PARI-ST       5/96       Ayyaz               
 A.~Babaev$^{24}$,                %ITEP-PD                  Babaev              
 J.~B\"ahr$^{35}$,                %ZEUT-PD                  Baehr               
 P.~Baranov$^{25}$
% ...

\end{flushleft}

\begin{flushleft} 
  {\it %     H1 Institutes as appearing on publications
\nopagebreak \noindent
% ...
$ ^2$ III. Physikalisches Institut der RWTH, Aachen, Germany$^a$ \\
% ...
$ ^{11}$ DESY, Hamburg, Germany$^a$ \\
% ...
$ ^{22}$ Department of Physics and Astronomy, 
          University of Manchester, Manchester, UK$^b$ \\
% ...
$ ^{24}$ Institute for Theoretical and Experimental Physics,
          Moscow, Russia \\
 $ ^{25}$ Lebedev Physical Institute, Moscow, Russia$^{f,k}$ \\
% ...
$ ^{28}$ LPNHE, \'{E}cole Polytechnique, IN2P3-CNRS, Palaiseau, France \\
 $ ^{29}$ LPNHE, Universit\'{e}s Paris VI and VII, IN2P3-CNRS,
          Paris, France \\
% ...
$ ^{34}$ Fachbereich Physik, Bergische Universit\"at Gesamthochschule
          Wuppertal, Wuppertal, Germany$^a$ \\
 $ ^{35}$ DESY, Institut f\"ur Hochenergiephysik, Zeuthen, Germany$^a$ \\
% ...
$ ^a$ Supported by the Bundesministerium f\"ur Bildung, Wissenschaft,
        Forschung und Technologie, FRG,
        under contract numbers 7AC17P, 7AC47P, 7DO55P, 7HH17I, 7HH27P,
        7HD17P, 7HD27P, 7KI17I, 6MP17I and 7WT87P \\
 $ ^b$ Supported by the UK Particle Physics and Astronomy Research
       Council, and formerly by the UK Science and Engineering Research
       Council \\
% ...
$ ^f$ Supported by the Deutsche Forschungsgemeinschaft \\
% ...
$ ^k$ Supported by Russian Foundation for Basic Research 
       grant no. 96-02-00019 
 }
\end{flushleft}
\clearpage
%================================================================

\section{Introduction}

The most precise determinations of the gluon momentum distribution  
in the proton have been obtained so far 
from the analysis of scaling violations
of the proton structure function 
%$F_2(x_B,Q^2)$~\cite{h1f2,zeusf2}
$F_2$~\cite{h1f2,zeusf2}.
This method is however indirect in the sense that
$F_2$ at low values of the Bjorken scaling variable $x_B$ 
actually probes the sea quark distributions
which are related via the QCD evolution equations to the gluon distribution.
The local behaviour of the structure function $F_2$ at a given value of $x_B$ 
depends on the gluon distribution $x_g g(x_g)$
in a rather wide range of values of the momentum fraction $x_g$, 
and the analysis 
requires the assumption of a certain functional form of $x_g g(x_g)$,
the parameters of which are then determined in a fit procedure. 

More direct determinations 
of the gluon density 
can be obtained by reconstruction 
of the kinematics of the interacting partons from the measurement of
the hadronic final state
in gluon-induced processes. 
Such direct measurements are complementary to the indirect ana\-ly\-ses:
although still limited in statistics, 
they are in principle more sensitive 
to local variations of the gluon distribution.
They are subject to different systematic effects and provide 
an independent test of perturbative QCD.
Direct gluon density determinations have previously been performed using 
events with $J/\Psi$ mesons in the final state~\cite{nmcjpsi}
and dijet events~\cite{dijet}.

The production of heavy quarks in electron-proton interactions
proceeds, in QCD, almost exclusively via photon-gluon fusion, where 
a photon coupling to the incoming electron interacts 
with a gluon in the proton by forming a quark-anti-quark pair.
This holds both for 
deep inelastic scattering (DIS)
and for
photoproduction where the exchanged photon is almost real.
Differential charm photoproduction cross sections in the
range of experimental acceptance 
were found to be reasonably well reproduced by such a description
in Next to Leading Order (NLO) QCD~\cite{dstargp},
and measurements of 
the charm contribution to the 
deep inelastic proton structure function, 
$F_2^c$~\cite{f2ch1z}, 
also confirm this picture.

Compared to the dijet case,  
smaller invariant masses of the   
partonic sub-process can be accessed in charm production. 
It is thus possible to extend the gluon density determination towards 
smaller momentum fractions $x_g$. 
The charm data are statistically less powerful, but there is also 
much less background from quark-induced processes.
 
Recently, Next-to-Leading-Order (NLO) QCD calculations of differential
charm cross sections at HERA have become 
available, both for DIS~\cite{bhmichi,bh111},  
and for 
the photoproduction re\-gime~\cite{frixione-1}.
With these calculations it becomes possible 
to determine $x_g g(x_g)$ directly.
Here, measurements of differential $D^{\ast}$ production 
cross sections with the H1 detector
are presented in both regimes, 
and from these results the gluon density is determined in NLO. 

The theoretical framework and the 
reconstruction of the gluon momentum fraction
will be described in more detail in the next section. 
In Section 3 
the measurement of differential $D^{\ast}$ production cross sections
will be presented separately for DIS and photoproduction.
In each regime, the results 
will be compared with predictions based on the appropriate 
NLO QCD calculations.
Section 4 will explain the method used to extract the gluon density
from the measured cross sections, and will present the results.
The paper concludes with a comparison to
results obtained by H1 from the analysis of 
the structure function $F_2$. 

%================================================================

\section{Principle of the Analysis}

% Theory input
\subsection{NLO QCD calculations}
Differential charm production cross sections in $ep$ interactions   
have been calculated~\cite{nason,bhcorrel} in NLO QCD
in the $\overline{\mbox{MS}}$ renormalization scheme.
The computations were carried out using the subtraction method, 
which is based on the replacement of divergent 
soft and collinear terms in the squared matrix elements 
by generalized distributions. 
Charm quarks are treated as massive particles appearing only in the 
final state whereas the three light 
flavours (and gluons) are the only active partons in the initial state
(``massive'' scheme, ``Three Flavour Number'' scheme).
In Leading Order, the production proceeds via the partonic sub-process
$\gamma g \ra c\bar{c} $. 
In NLO, there are in addition contributions to the cross section
associated with diagrams 
where a real gluon is radiated from a gluon or a quark line, 
$\gamma g \ra c\bar{c} g$, 
there are quark induced contributions via the processes
$\gamma q  \ra c\bar{c}  q$ and 
$\gamma \bar{q} \ra c\bar{c}\bar{q}$,
and there are virtual corrections to the cross section.  

For the photoproduction processes, resolved photon--proton interactions
have to be taken into account in addition to the direct photon--proton
scattering. While in the latter processes, the photon couples directly to
a parton of the proton, the resolved photon processes involve the partonic
structure of the photon itself.
In NLO, the distinction of these processes is not unambiguous; 
here, a part of the photon parton density is contained in the direct
photon calculation. 
The remaining part 
that depends on the parton 
density functions of the photon is termed resolved. 
In the calculations of charm production performed in the `massive' scheme 
the direct contribution is dominant 
in the phase space where at least one charmed
meson is produced in the detector acceptance.
Previous measurements of charm photoproduction~\cite{dstargp}
have confirmed this. 

Depending on the kinematic regime under consideration,  
the QCD scales in the theoretical calculations 
are governed by different quantities.
In the DIS case the factorization scale is set to be
\mbox{$\mu_F = \sqrt{4m_c^2+Q^2}$},
where $m_c$ denotes the charm quark mass and 
$Q^2$ the virtuality of the exchanged photon,
and the renormalization scale is set to be $\mu_R=\mu_F$.
In the calculation of $\gamma p$ cross sections, 
the factorization scale is given by 
the transverse momenta of the charm quarks: 
\mbox{$\mu_F = \sqrt{4m_c^2+4p_{\perp}^2}$},
and the renormalization scale is chosen to be 
$\mu_R=\mu_F/2\,$.
Variation of these choices will be one means 
of estimating the uncertainties of the perturbative calculations.

Since the computations are differential in all relevant quantities,
they provide access to the four-momenta of the outgoing partons.
The evaluation of cross sections can be performed 
using Monte Carlo integrators. 
The calculations by Harris and Smith for DIS have been implemented in the 
HVQDIS program~\cite{bh111}, and those 
for the photoproduction case 
by Frixione {\it et al.}\
in the FMNR package~\cite{frixione-1}.  

Charm quark hadronization into $D^{\ast}$ mesons
is performed in the programs 
by convoluting the charm quark cross section 
with the Peterson fragmentation function~\cite{peterson};
the $D^{\ast}$ three-momentum vector is formed by scaling the quark 
three-momentum vector, and the $D^{\ast}$ energy is fixed such that the mass is 
$m(D^{\ast})=2.01$ GeV. 
Evolution of the fragmentation function,
which one expects to become important 
for transverse $c$ quark momenta $p_{\perp}\gg m_c$,
is not included because with present experimental cuts
the region of interest is $p_{\perp}\sim m_c$.
Cross sections for charm photoproduction at 
large transverse momenta $p_{\perp}\gg m_c$
have been computed 
in the``massless'' scheme~\cite{kniehl,kniehlhera}, 
where charm acts as an active flavour in the initial state proton and photon. 

Using the HVQDIS ~\cite{bh111} and FMNR~\cite{frixione-1} programs, 
cross sections incorporating experimental cuts
can be calculated in NLO as a function of any final state variable 
depending on the $D^{\ast}$ and electron four-momenta. 
Thus the predictions of the NLO QCD
calculations can be confronted directly with measured differential
cross-sections for inclusive $D^{\ast}$ production in the 
experimentally accessible kinematic range.
 
% x_gluon reconstruction
\subsection{Reconstruction of the gluon momentum fraction}
\label{sec:recx}
The kinematics of the Leading Order $2\ra 2$ process 
$\gamma g \ra c\bar{c}$
is completely determined if the momenta of 
one incoming and one outgoing particle, here the photon and the 
charm (anti-) quark, 
are known.
The momentum fraction $x_g$ 
in the infinite momentum frame is given by 
\begin{equation}\label{xgkine}
%x_g = x_B \left ( 1 + \frac{M^2}{Q^2} \right )
x_g = \frac{M^2+Q^2}{y\cdot s}
\;\;\; \mbox{with} \; \;\;
M^2 = \frac{p_{\perp c}^{*2} + m_c^2}{z(1-z)}   
\;\;\; \mbox{and} \;\;\; 
z \equiv \frac{p\cdot p_c}{p\cdot q} = \; 
\frac{(E-p_z)_c^{\mbox{\tiny (Lab)}}}{2yE_e} \;\; ,
\end{equation}
where $p_{\perp c}^*$ is the transverse momentum of the 
charm quark 
with respect to the photon-proton axis
in the photon-proton centre-of-mass frame. 
The Lorentz invariant $z$ 
can be evaluated in the laboratory frame
from the charm quark's energy $E$ and its momentum component $p_z$.%
\footnote{H1 uses a right-handed coordinate system with the $z$ 
axis defined by the incident proton beam, and the $y$ axis pointing upward.}
The variables $p$, $q$, and $p_c$ denote the four-momenta of 
the incoming proton, exchanged photon,  
and outgoing (anti-)charm quark, respectively, 
so that the usual DIS variables read
$Q^2=-q^2$ and
%and $x_B=Q^2/(2p\cdot q)$ the Bjorken scaling variable; 
%$y=Q^2/x_Bs$ , 
$y=(2p\cdot q)/s$; 
$\sqrt{s}$ is the $ep$ centre of mass energy, 
and $E_e$ the incoming electron energy in HERA.
In DIS, Eq.~\ref{xgkine} can be written in the form 
$x_g = x_B ( 1 + M^2/Q^2)$,
which relates $x_g$ to the Bjorken scaling variable $x_B$.  
 
In the presence of gluon radiation (or a gluon transverse momentum 
in the initial state),
relation~(\ref{xgkine}) holds
only approximately. 
Also, in the experiment one measures not quarks 
but $D^{\ast}$ mesons that on average carry only a fraction of the 
energy of the primarily produced $c$ quark. 
However, 
one can construct a hadronic observable $x_g^{OBS}$
that is well correlated with the true $x_g$ 
by replacing in Eq.~\ref{xgkine}
\begin{equation}\label{repl}
p_{\perp c}^*  \ra  1.2 \, p_{\perp D^{\ast}}^* \; , \;\;\;  
(E-p_z)_c  \ra   (E-p_z)_{D^{\ast}} \;\; .
\end{equation}
The factor 1.2 is introduced for convenience in the 
unfolding procedure described in Sect.~\ref{sec:glue}; it 
is chosen such that $x_g^{OBS}\approx x_g$ on average. 

\subsection{Analysis procedure}
\label{sec:proc}
% Structure of analysis 
The analysis proceeds in two main steps.
First, the $D^{\ast}$ cross section is measured 
in the region of experimental acceptance
as a function of various kinematic variables and
of $x_g^{OBS}$. 
The data are corrected 
only for detector effects. 
As there is no extrapolation into unmeasured regions of phase space
nor are there any parton-hadron correlations involved, 
this step can be performed using the Leading Order
Monte Carlo programs which describe  
the final state reasonably well.
The resulting cross sections are then confronted with the NLO QCD predictions.

In the second step the cross section 
as a function of $x_g^{OBS}$ is unfolded 
to the true variable $x_g$ and the gluon density $x_g g(x_g,\mu_F^2)$ 
is extracted.
This step is carried out in NLO QCD in the framework 
of the Three Flavour Number Scheme.
The programs HVQDIS for DIS and FMNR for photoproduction 
are used to
calculate the cross sections 
in the region of acceptance
as a function of $x_g^{OBS}$ 
as well as the correlations between 
the hadronic and partonic quantities, $x_g^{OBS}$ and $x_g$,
taking into account 
NLO effects and fragmentation.

%================================================================

\section{Cross Section Measurement}

%HERA
\subsection{Experimental conditions}
\label{sec:excon}
The data used here have been collected 
in the years 1994 to 1996 with the H1 detector 
at the HERA collider, 
where 27.5\,\gev\ positrons
(henceforth generically termed ``electrons") 
collided with 820\,\gev\ protons,
at a centre of mass energy of $\sqrt{s}=$ 300\,\gev.

%H1 detector
The H1 detector and its trigger capabilities have been 
described in detail elsewhere~\cite{H1det}.
Charged particles are 
measured by two cylindrical jet drift chambers \cite{cjc,cjc-res}, 
mounted concentrically around the
beam-line inside a homogeneous magnetic field of 1.15 Tesla,
yielding particle charge and momentum from the track curvature
in the polar angular range
of 20$^{\circ}<\theta<160^{\circ}$,
where $\theta$ is measured
using the central jet and
two $z$ drift chambers and is defined with respect to the 
incident proton beam direction. 
One double layer of cylindrical multi-wire proportional 
chambers (MWPC) \cite{mwpc} with pad readout
for triggering purposes is positioned inside and another one  
in between the two jet chambers.
The tracking detector is surrounded by a fine grained liquid
argon calorimeter \cite{calo}, consisting of an electromagnetic
section with lead absorbers and a hadronic section with
steel absorbers.
It covers polar angles between 4$^{\circ}$ and 154$^{\circ}$.
In the backward region ($153^{\circ}<\theta<177.8^{\circ}$) 
it is complemented by 
a lead scintillator ``Spaghetti" calorimeter (SpaCal)~\cite{spacal}
which is optimized for the detection of the scattered electron in the 
DIS kinematic range under consideration here
and provides Time-of-Flight functionality 
for trigger purposes. 
Installed in 1995, 
it consists of an electromagnetic and a more coarsely 
segmented hadronic section.
A four-layer drift chamber (BDC)~\cite{bdc}
mounted on its front improves
the angular measurement of the scattered electron.

The luminosity is determined from the rate of {\em Bethe-Heitler}
$e p \rightarrow e p \gamma$ bremsstrahlung events.
The luminosity system
consists of an electron detector and a photon detector,
located 33~m and 103~m from the interaction point
in the electron beam direction, respectively.
The system is also used to tag photoproduction
events by detecting electrons scattered at small angles.
In 1995 and 1996 a second electron tagging detector of the same
type was added at 44 m. 
It covers a much lower $y$-range.
Therefore the two electron detectors cover different 
regimes of photon-proton centre-of-mass energies, $W_{\gamma p}$,
and thus are complementary.
Time-of-flight (TOF) systems in the forward and backward
directions are used to reject beam gas background.

%================================================================

% Common DIS and gp issues
\subsection{Extraction of the \boldmath $D^{\ast}$ \unboldmath signal }
\label{sec:dstsig}
\begin{figure}[tbp]\centering
\unitlength1.0cm
\begin{picture}(17,20)(0,0.4)
\put(3.,13.6){\epsfig{file=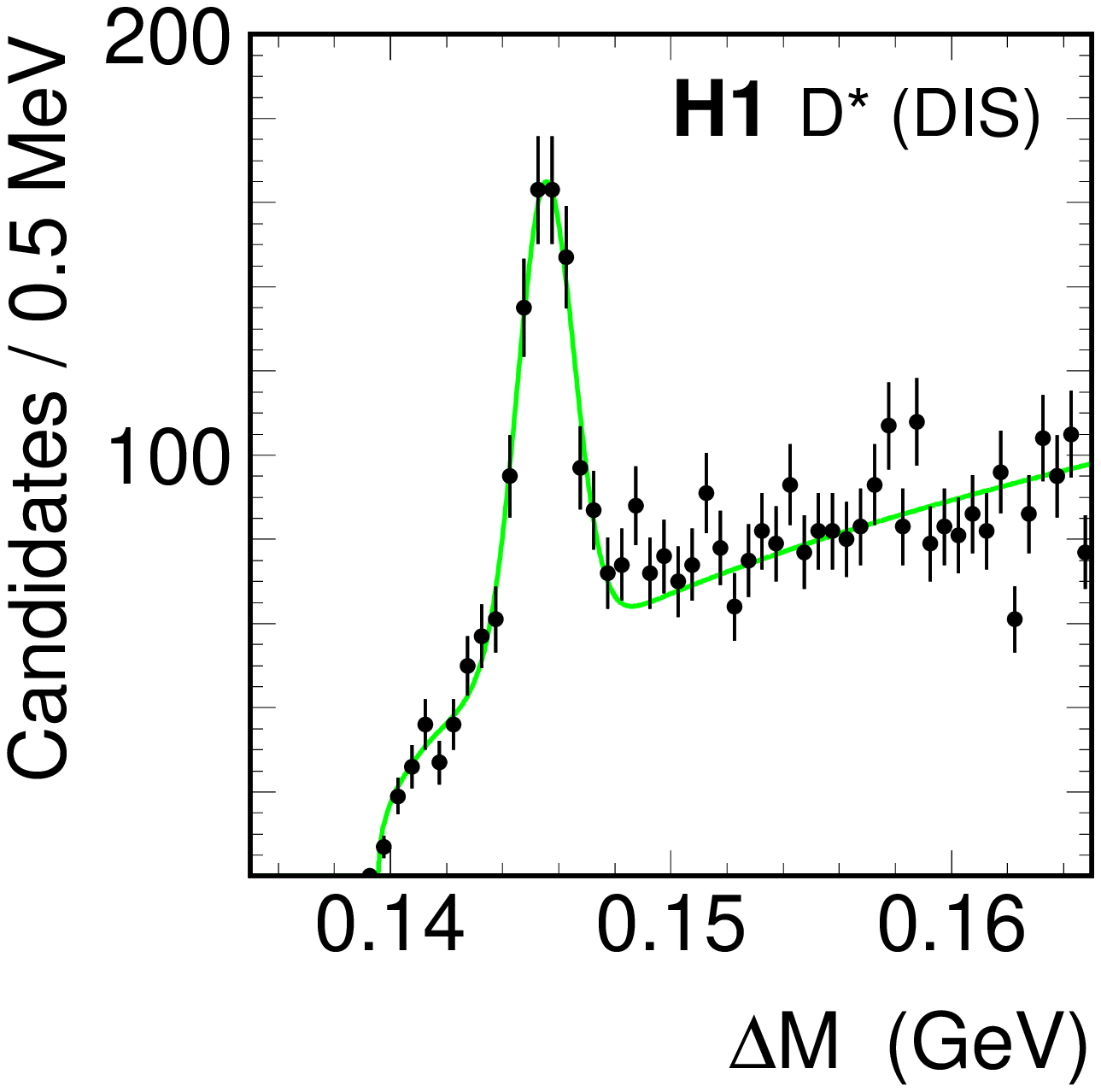,width=8cm}}
\put(3.,6.8){\epsfig{file=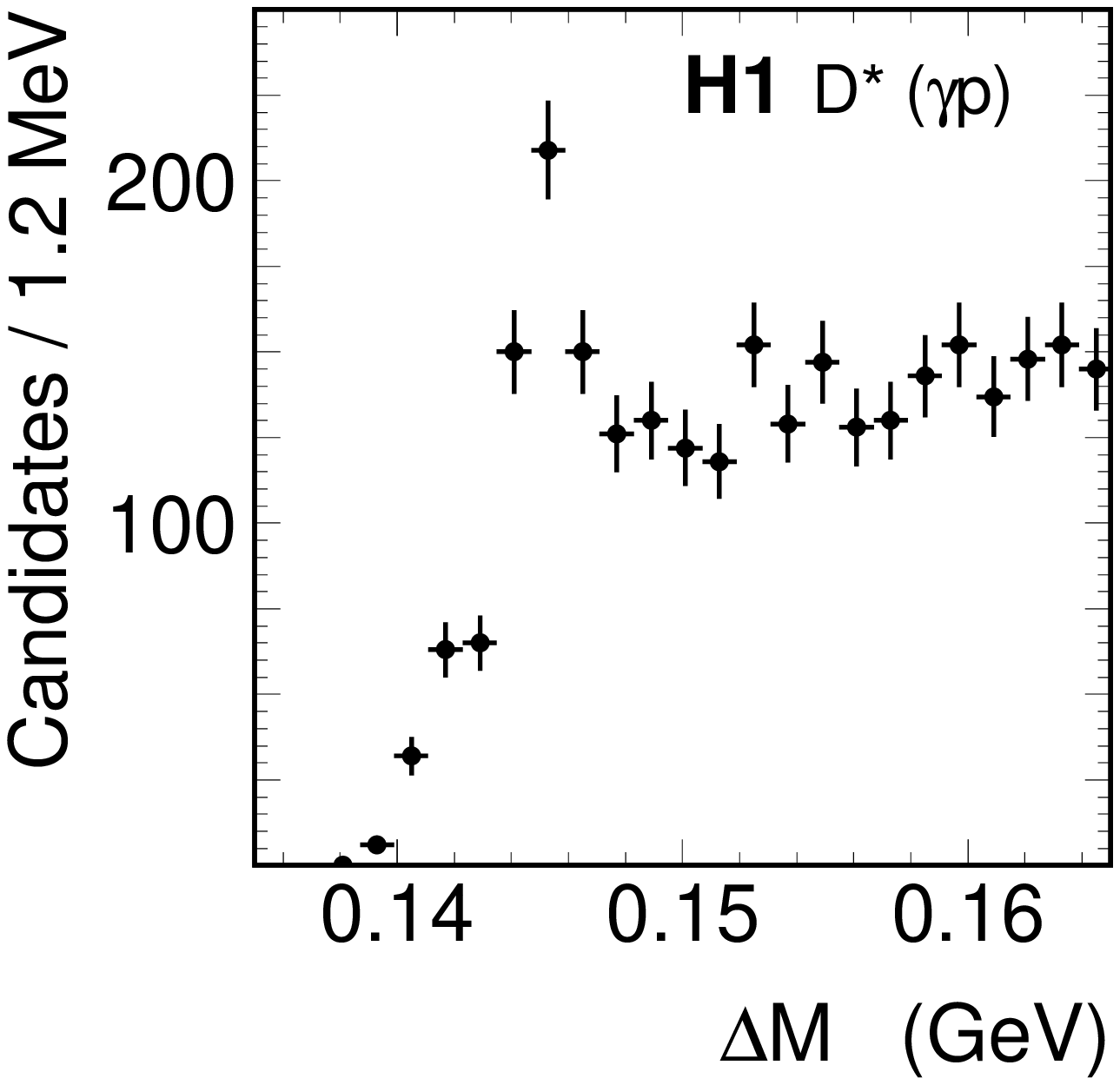,width=8cm}}
\put(3.,0.){\epsfig{file=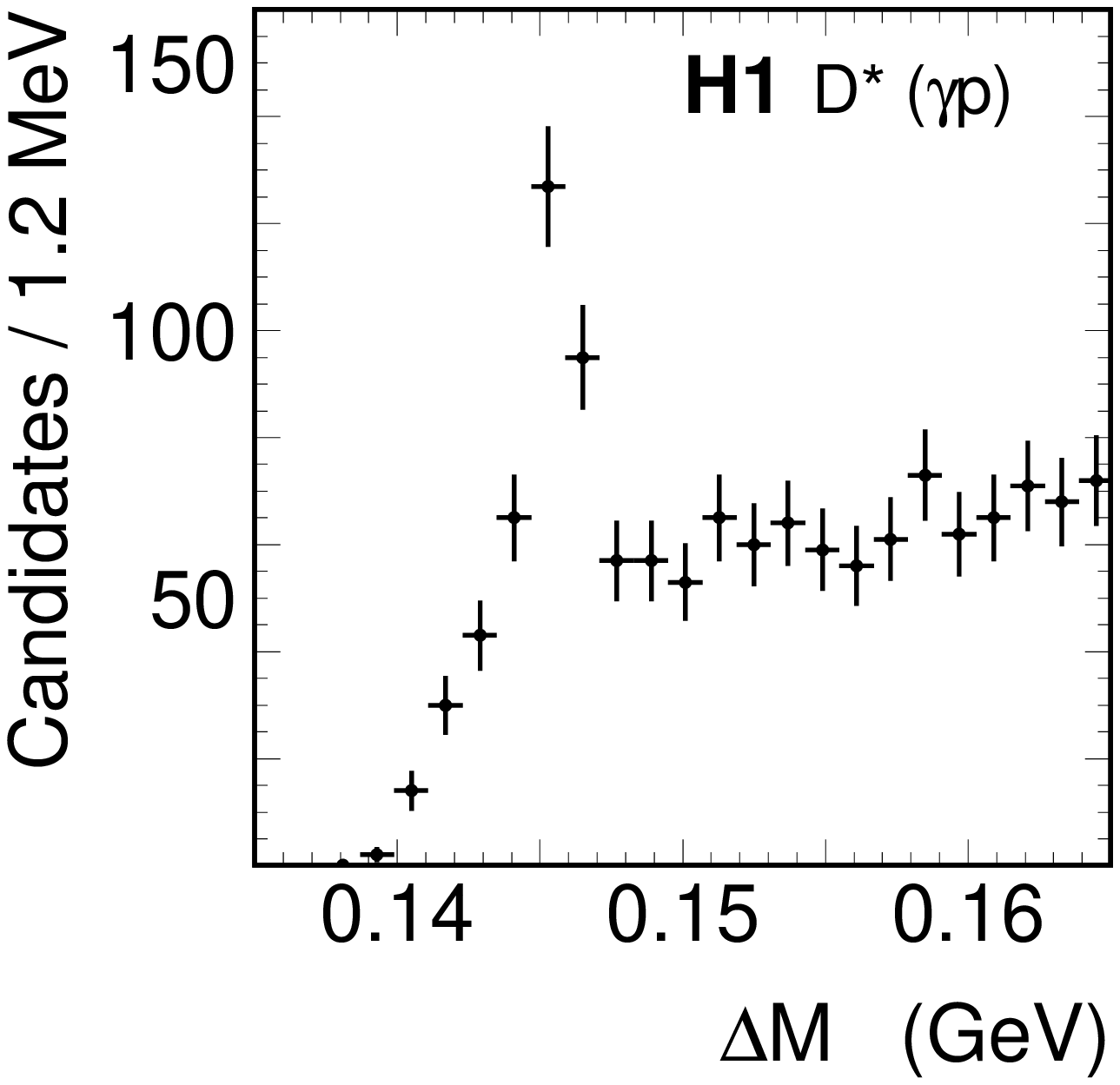,width=8cm}}
\put(5.3,18.6){a)}
\put(5.3,11.8){b)}
\put(5.3,5.0){c)}
%\put(0,0){.(0,0)}
\end{picture}
\caption[dummy]{\label{fig:Dstarpeak}\sl%
         (a) Mass difference 
         $\Delta M  = M (K^-\pi^+ \pi^+ _{slow}) - M (K^-\pi^+)$ 
         distribution 
         of $D^{\ast}$ candidates
         in DIS.  
         The solid line represents the result of 
         a fit as described in the text,
         which is used to extract the $D^{\ast}$ cross section. 
         For the photoproduction samples
         {\it ETAG33} (b) and {\it ETAG44} (c)
         the unweighted
         $\Delta M$ distributions are shown. 
} 
\end{figure}

The reconstruction of $D^{\ast}$ mesons in the DIS and photoproduction regime
follows closely 
the method described
in the previously published measurement of $D^{\ast}$ 
production~\cite{dstargp}.
It makes use of the $D^{\ast}$ tagging technique~\cite{feldmann},
{\it i.e.}\ of the tight kinematical conditions
in the decay chain 
$D^{\ast +} \ra D^0\pi^+ _{slow} \ra (K^-\pi^+) \pi^+_{slow}$%
\footnote{The charge conjugate is always implied.}
for which the overall branching fraction is
2.62\%~\cite{pdg}.
Particle tracks passing some quality cuts
and fulfilling 
transverse momentum cuts depending on the kinematic range, 
are first combined in unlike-sign charged pairs.
No particle identification is applied. 
Assuming one track to  
be a kaon and the other to be a pion,
the invariant mass is calculated for all combinations 
and each possible mass hypothesis assignment.  
Those $K^-\pi^+$ pairs with an invariant mass 
consistent with the $D^0$ mass within $\pm 80$ MeV
are combined with a third track (``$\pi^+_{slow}$'')
which has to have charge of opposite sign to that of
the kaon candidate and to which is assigned the pion mass hypothesis.
$D^{\ast +}$ production is found as a distinct enhancement 
in the distributions of the mass difference
$\Delta M  = M (K^-\pi^+ \pi^+ _{slow}) - M (K^-\pi^+)$
around the expected mass difference of $145.4$\,MeV~\cite{pdg}.
The $\Delta M$ distributions for all $D^{\ast}$ candidates 
are shown in
Fig.~\ref{fig:Dstarpeak}
for the DIS (a) and the two photoproduction samples (b,c).
No enhancement is observed if the mass difference for the
wrong charge combinations 
$ M(K^- \pi^- \pi^+) - M(K^- \pi^-)$
is used instead.

The number of events, $N_{D^{\ast}}$, is extracted from 
maximum likelihood fits to 
the distributions of the mass difference
%$\Delta M  = M (K^-\pi^+ \pi^+ _{slow}) - M (K^-\pi^+)$
with a function taken to be a superposition of a Gaussian for the signal
and a term $N_B (\Delta M-m_{\pi})^a$ for the background. 
The position and width of the Gaussian are fixed to the values
found from the total sample including both 
photoproduction and DIS events.
The position is consistent with the expected value, 
and the width agrees with simulation results.
The normalization of the signal,
the background normalization $N_B$, and the shape exponent $a$ are free
parameters in the fit. 
%Other methods of determining the number
%of candidates have also been applied 
%and yield the same  numbers
%within the statistical errors.
Uncertainties from 
the parameterization of signal and 
background
are accounted for in the systematic error.
From Monte Carlo simulations the contribution
to the signal 
%in DIS 
due to $D^0$ decays into channels other than $K^-\pi ^+$
(mostly \mbox{$D^0\ra K^-K^+$}
and \mbox{$D^0\ra \pi^-\pi^+$})  
is obtained to be $r=(3.5\pm 1.5)$\%.
For these reflections a correction is made. 

\subsection{Acceptance and efficiency determination}
A Monte Carlo simulation  
is used to determine the detector acceptance
and the efficiency of the 
reconstruction and the
selection cuts.  
Electroproduction events and
direct and resolved
photoproduction events were generated in Leading Order with the
AROMA 2.2~\cite{aroma}
and PYTHIA 5.7 programs~\cite{pythia}.
The hadronization step was performed according to the 
Lund string model.
The generated events were then
processed by
the H1 detector simulation program,
and were subjected to the same reconstruction and analysis
chain as the real data.
The observed rapidity and transverse momentum distributions 
of $D^{\ast}$ mesons are acceptably reproduced.
The changes found when varying the shape of 
the charm fragmentation function by using a Peterson 
function with $\epsilon_c$ ranging from 0.026 to 0.046
are accounted for in the systematic error. 
The dependence of the simulated efficiencies on other 
parameter choices
made for the simulation (charm mass, parton density distributions) 
was studied, found to be small, and is neglected. 

Contributions from $b\bar{b}$ production, with subsequent decays
of $b$ flavoured hadrons into $D^{\ast}$ mesons, have been calculated 
using the AROMA generator 
to be about 1.6\% for DIS and even smaller for photoproduction. 
No subtraction is made;
the quoted $D^{\ast}$ cross sections are inclusive. 
A systematic error
accounts for the change of the efficiency due to 
a variation of the $b\bar{b}$ contribution within a factor of 5
of its predicted value, thus taking the preliminary result~\cite{h1bprel}
into account, where an unexpectedly high $b\bar{b}$ cross section 
at HERA was found.

\subsection{Differential cross sections}

Differential cross sections as functions of kinematic variables 
are measured by dividing the data sample into bins 
and 
determining the numbers of
$D^{\ast}$ mesons and the efficiencies in each bin 
using the $\Delta m$ fit and correction procedure
as outlined above.
The effects of limited detector resolution on the 
variables are small  
with respect to the chosen bin sizes 
and are corrected for using the Monte Carlo simulation. 

The cross sections 
are determined in the ranges
specified in Tab.~\ref{tab:sigranges},
\begin{table}[bht] \centering
\caption{\label{tab:sigranges}\sl
Kinematic ranges of the cross section measurements.
}
\vspace{5mm}
\begin{tabular}{|c||c|c|}
\hline
& & \\ [-3.5mm]
DIS & $\gamma p$, {\it ETAG44} & $\gamma p$, {\it ETAG33} \\[1mm] 
\hline
& & \\ [-3.5mm]
$2 < Q^2 < 100$ GeV$^2$ &
$Q^2 < 0.009$ GeV$^2$   & $Q^2 < 0.01$ GeV$^2$ \\
$0.05 < y < 0.7$        & 
$0.02 < y < 0.32$        & $0.29 < y < 0.62 $ \\
$p_{\perp}(D^{\ast}) > 1.5$ GeV &
 $p_{\perp}(D^{\ast}) > 2$ GeV & $p_{\perp}(D^{\ast}) > 2.5$ GeV \\
$|\eta (D^{\ast})| < 1.5$ & $|\hat{y} (D^{\ast})| < 1.5$ & $|\hat{y} (D^{\ast})| < 1.5$  \\ [1mm]
\hline
\end{tabular}
\end{table}
where $\eta = - \ln \tan (\theta /2)$ denotes the pseudo-rapidity
in the laboratory frame. 
In photoproduction, where the photon-proton 
centre-of-mass frame moves along the $z$ axis, the rapidity 
in the laboratory frame,
\mbox{$\hat{y} =  - \frac{1}{2} \ln{\frac{E - p_z}{E + p_z}}$}, 
is conveniently used. 

In the following, 
the DIS and photoproduction analyses are presented separately,
since the scattered electron is measured in different detectors 
%-- in the SpaCal and in the electron taggers, respectively --  
and since different methods are used to reconstruct the event kinematics. 
Different cuts in 
the $D^{\ast}$ selection reflect different combinatorial background levels
in the kinematic regimes under consideration. 

%================================================================

%DIS 
\subsubsection{Deep Inelastic Scattering analysis}\label{sec:disana} 
  
The data have been taken in the years 1995-96 after the 
installation of the SpaCal and the BDC. 
They correspond to a total integrated luminosity of 
${\cal L}  = 9.7$ pb$^{-1}$.
The events have been triggered by an electromagnetic cluster
in the SpaCal
of at least 6.5 GeV energy, 
in coincidence with a charged track signal
from the MWPC and central drift chamber trigger.

The identification of electrons is similar to the procedure
used in the inclusive structure function measurement~\cite{f295}.
Scattered electrons are identified as clusters in the SpaCal with 
energy $E_{e^{\prime}}>8$ GeV, with a cluster radius $<$3.5 cm consistent with
electromagnetic energy deposition, and with 
a cluster centre matched 
by a charged track candidate
in the BDC within 2.5 cm. 
The scattering angle is required to be 
$\theta_{e^{\prime}}<177^{\small o}$. 

The DIS kinematic variables are 
reconstructed using the ``electron ($e$)'' 
and the ``$e\Sigma$'' me\-thod~\cite{bbkine}.
In both methods 
\begin{equation}
Q^2  = 4 E_e E_{e^{\prime}} \cos ^2 \frac{\theta_{e^{\prime}}}{2} 
\end{equation}
whereas 
\begin{equation}
y_e = 1 - (E_{e^{\prime}}/E_e) \sin ^2 (\theta_{e^{\prime}} /2)\;\; 
\;\;\; \mbox{and} \; \; \; 
y_{e\Sigma} = \frac{2 E_e \sum_{had} (E-p_z)}
         {\left[\sum_{all} (E-p_z) \right ]^2}         \;\; ,
\end{equation}
where the sum in the denominator 
runs over all 
(liquid Argon and SpaCal) calorimetric signals as well as 
central drift chamber tracks, avoiding double counting. 
The sum in the numerator is over 
the same objects
except those 
forming the scattered electron candidate. 
The results obtained with both methods are found to be 
in very good agreement. 
The  $e\Sigma$ method needs smaller QED radiative corrections,
and it is less sensitive to uncertainties in the 
electron energy. 
The absolute energy scale of the SpaCal was known to a precision 
of $\pm 2\%$ (at 27 GeV) at the beginning of its operation in 1995, 
and to $\pm 1\%$ in 1996. 
The $e\Sigma$ method is therefore used for the 1995 data.
For the 1996 data, the $e$ method is used,
which yields slightly improved average resolution in
$x_g^{OBS}$.

For $D^{\ast}$ reconstruction, cuts 
of \mbox{$p_{\perp}> 300\,$ (120) MeV} are imposed on the 
on the transverse momenta in the laboratory frame 
$K$ and $\pi$ ($\pi_{slow}$) candidates. 
In order to suppress combinatorial background,
the range of $z$ (Eqs.~\ref{xgkine},\ref{repl}) 
is restricted to $z>0.2$ for $p_{\perp}(D^{\ast})<3$ GeV. 
Variations found by applying or not applying this cut are accounted for
in the systematic error. 

A total of $N_{D^{\ast}} = 583 \pm 35$ events is observed;
a correction is made for 
a small fraction $r$ which is attributed to reflections
(see Sect.~\ref{sec:dstsig}).
The efficiency for electron and $D^{\ast}$ reconstruction and selection 
is found to be 
$\epsilon_{reco}=42\%$ on average, 
% with a moderate dependence on $x_g^{OBS}$, 
ranging from
34\% in the lowest to 57\% in the highest $x_g^{OBS}$ bin. 
The trigger efficiency is determined from data, by
using independent trigger conditions.
It is found to be $\epsilon_{trig}=98.6\%$ 
on average for the 1995 running period,
and 83.3\% for 1996 when tighter track conditions were applied.  
The dependence of $\epsilon_{trig}$
on the event kinematics is small and is taken from 
simulations. 
 
There is no indication of photoproduction background
seen in the data; the electron energy and $y_{\Sigma}$ distributions 
are consistent with expectations for a pure sample of DIS events.  
From simulations of direct
and resolved charm photoproduction, 
we expect 
less than 1 event within the cuts specified.
No subtraction is therefore made and a 1\% error attributed to
this possible background source. 

Radiative corrections have been calculated using the program 
HECTOR~\cite{hector}. 
They amount to $\delta _{rad}=$ 3\% (7\%) on average 
for the $e\Sigma$ ($e$) method and are below 10\% in all bins.  
The cross section for the given range in $Q^2$, $y$, $p_{\perp}$ and $\eta$
(Tab.~\ref{tab:sigranges}) is obtained as
\begin{equation}\label{eq:xect}
\sigma = \frac{N_{D^{\ast}}(1-r)} 
            {\epsilon_{trig} \epsilon_{rec} \, 
            BR(D^{\ast}\ra K\pi\pi )\,  {\cal L}\, (1+\delta_{rad})} 
% = (5.63 \pm 0.66 \pm 0.81 ){\rm nb}\; , 
%  = (5.63 \pm 0.66^{+0.84}_{-0.68}){\rm nb}\; , 
%  = (5.62 \pm 0.33 \pm 0.95){\rm nb}\; , 
%   = (5.54 \pm 0.34 \pm 0.94){\rm nb}\; , 
   = (5.75 \pm 0.35 \pm 0.79){\rm nb}\; , 
\end{equation} 
where the first error is statistical, the second is systematic
and is detailed in
Tab.~\ref{tab:sigsyst}.
\begin{table}[bht] \centering
\caption{\label{tab:sigsyst}\sl
Systematic errors of the cross section measurements.
}
\vspace{5mm}
\begin{tabular}{|l|r|r|}
\hline
%\multicolumn{4}{|c|}{Applied Corrections}\\
%\hline
   & DIS & $\gamma p$ \\
\hline
Trigger efficiency & +4.3, -3.3\% & \hspace{12mm}6.5\% \\
\hline
Track reconstruction  & \multicolumn{2}{c|}{7.5\%} \\
\hline
$D^{\star}$-Fit &  +3.7, -4.4\% & 6\% \\
\hline
Reflections & \multicolumn{2}{c|}{1.5\%} \\
\hline
$b$-production background & \multicolumn{2}{c|}{1\%} \\
\hline
$\gamma p$ background & +0, -1\% & -- \\
\hline
Radiative corrections &  2.9\%  & --  \\
\hline 
Energy scale elm. SpaCal & 3.3\% & --\\
%1\% at 27GeV, 3\% at 7GeV & 3.3\% & -- \\
\hline
Energy scale Had SpaCal ($\pm$7\%) & $<$  1\% & -- \\
\hline
Energy scale LAr ($\pm$ 4\%) & $<$ 1\% & -- \\
\hline
Electron angle ($\pm$0.5 mrad) & $<$ 1\% & -- \\
\hline
BDC hit finding & 2\% & -- \\
\hline
Electron tagger acceptance & -- & 6\%\\
\hline
Luminosity & \multicolumn{2}{c|}{1.5\%} \\
\hline
branching fraction & \multicolumn{2}{c|}{4\%} \\
\hline
z - cut & +7.5, -6.6\% & -- \\
\hline
Fragmentation & +0.6\%, -2.4\% & 3\% \\
\hline
\hline
Total & +14.1, -13.4 \% & 15\% \\
\hline
\hline
\end{tabular}
\end{table}
One of the largest contributions in both analyses is due to the uncertainty
in the track reconstruction efficiency.
Other important sources of systematic errors include 
uncertainties in the trigger efficiency and 
in the $D^{\ast}$ signal extraction.
In DIS, uncertainties related to  
the modelling of the $D^{\ast}$ distributions, 
as reflected in the dependence on the $z$ cut,
are also sizable. 
All contributions show only slight variation with $x_g^{OBS}$. 

\begin{figure}[tbp]\centering
\unitlength1cm
\begin{picture}(17,14)
\put(0.0,7.6){\epsfig{file=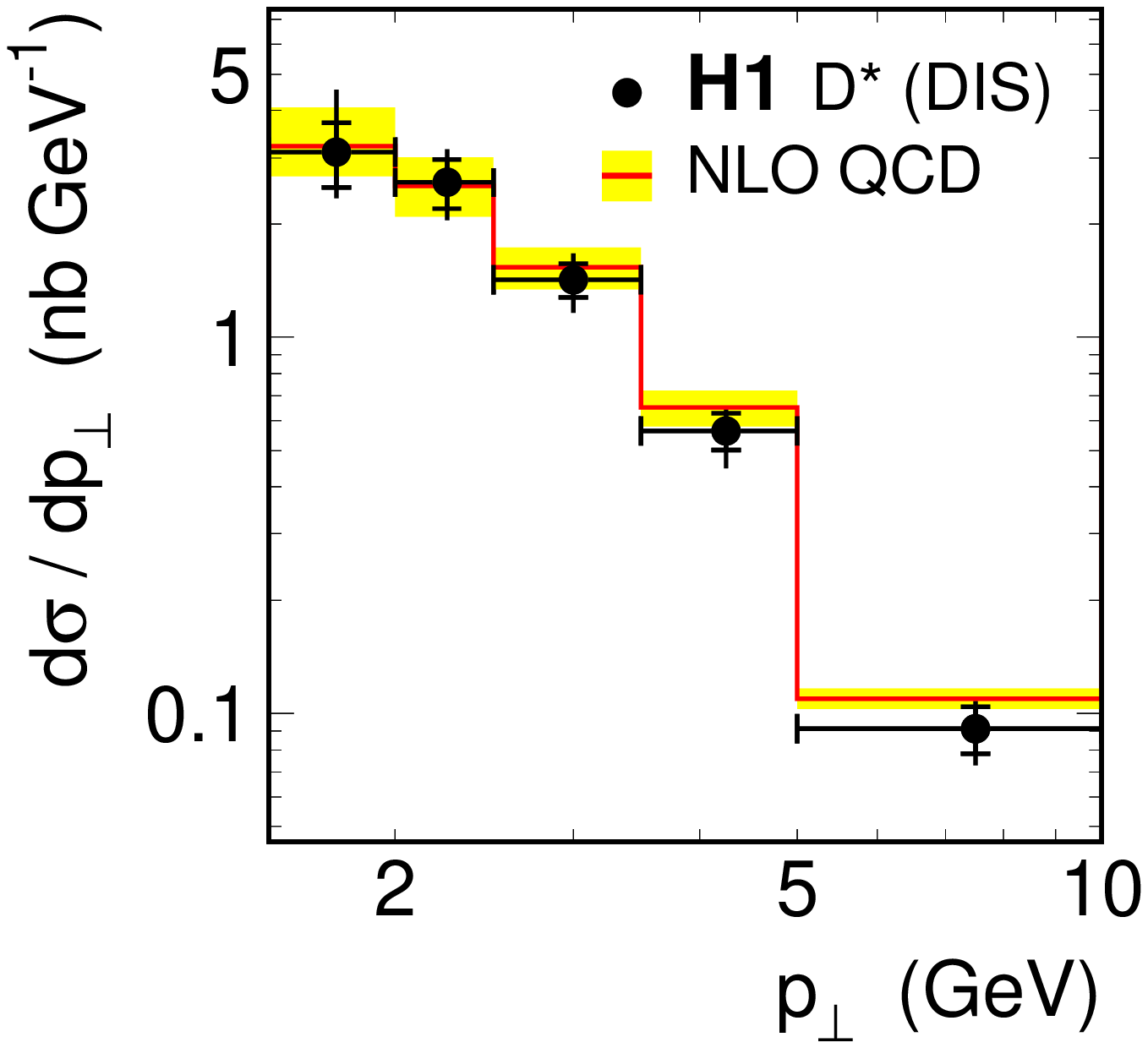,width=8cm}}
\put(8.,7.5){\epsfig{file=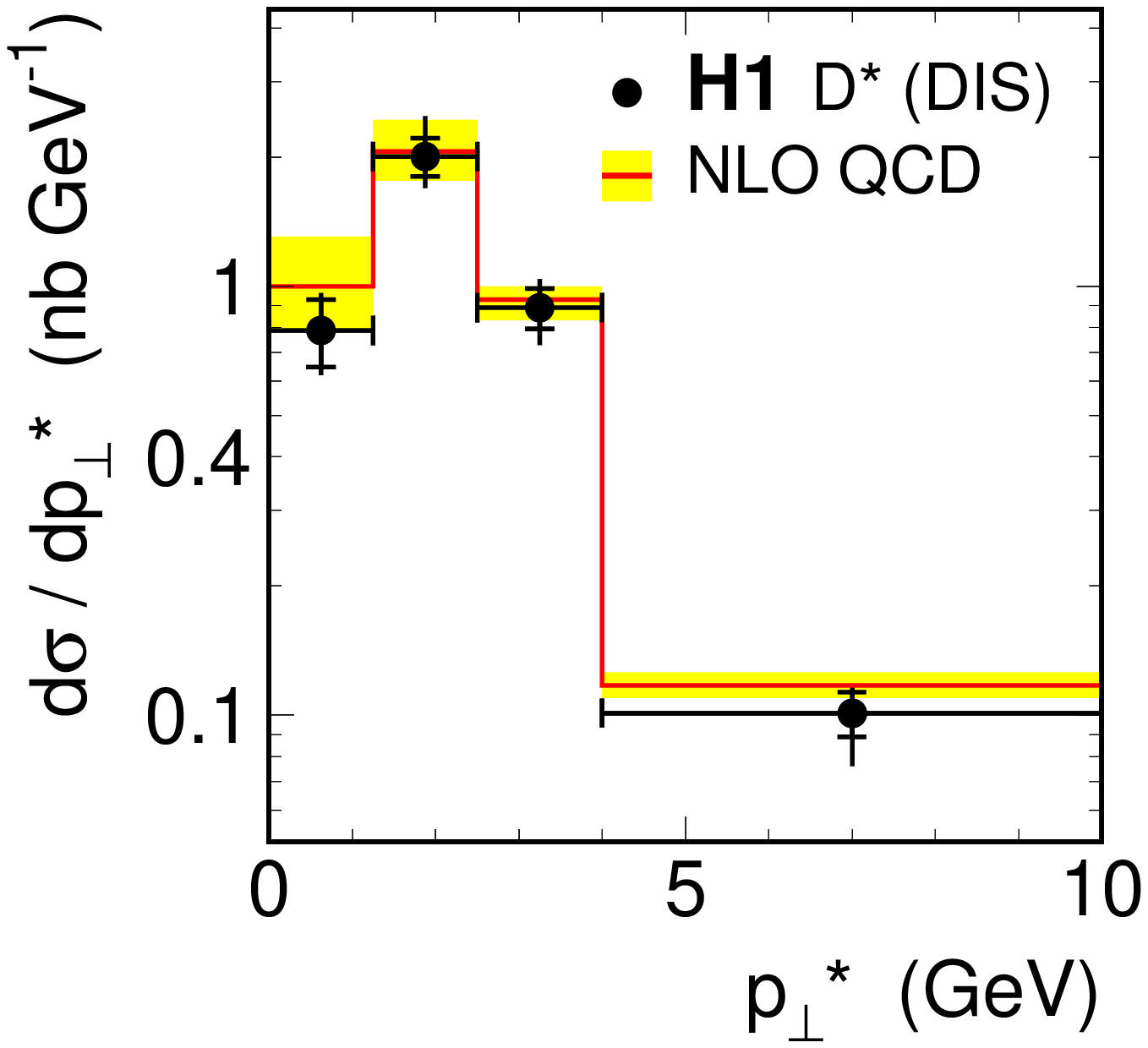,width=8cm}}
\put(0.0,0.0){\epsfig{file=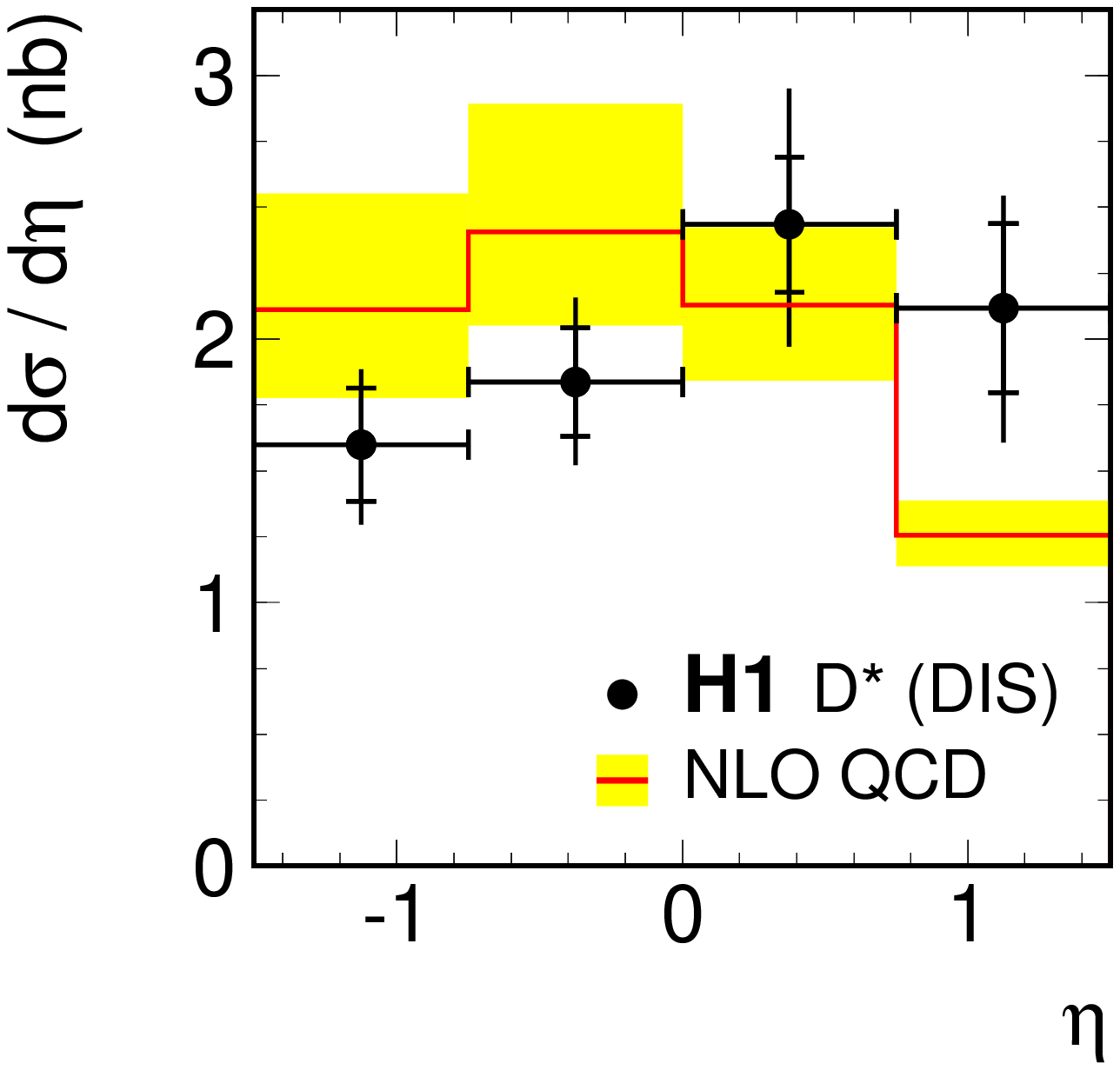,width=8cm}}
\put(8.,0.0){\epsfig{file=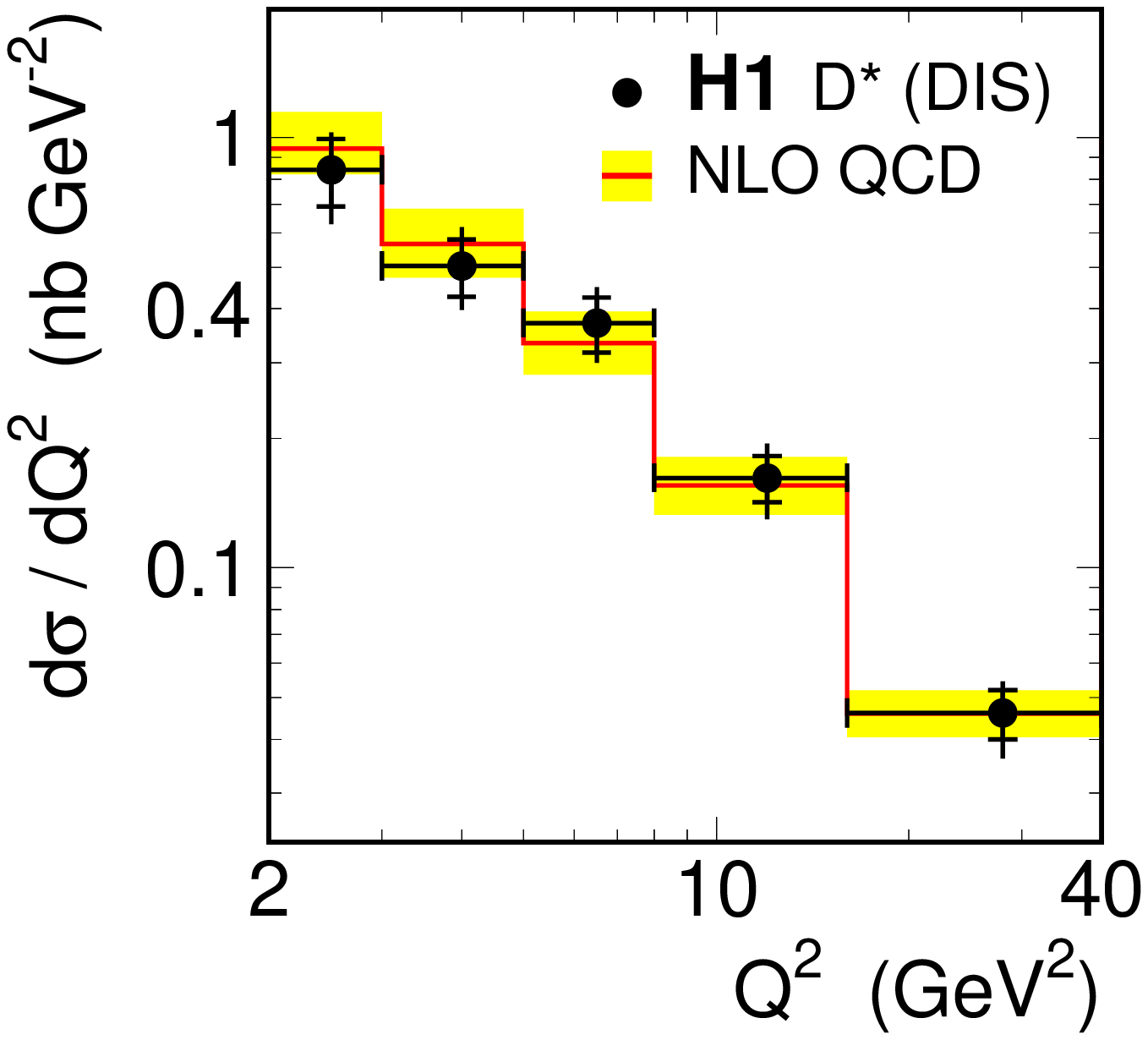,width=8cm}}
\put(2.3,9.5){a)}
\put(10.3,9.5){b)}
\put(2.3,2.){c)}
\put(10.3,2.){d)}
%\put(0,0){.(0,0)}
\end{picture}
\caption[dummy]{\label{fig:difxsect}\sl%
         Differential DIS cross sections in
         the kinematic range of experimental acceptance
         (see Tab.~\ref{tab:sigranges}). 
         The H1 data are shown as points with error bars
         (inner: statistical, outer: total); 
         the NLO QCD prediction 
         using the CTEQ4F3 parton distribution set 
         and a charm mass $m_c=1.5$~GeV 
         is shown as a histogram.
         The shaded band represents the variation of the 
         theoretical cross sections 
         due to different choices of $m_c$ between 1.3 and 1.7 GeV.
         (a) Transverse $D^{\ast}$ momentum in the laboratory frame,
         (b) Transverse $D^{\ast}$ momentum in the hadronic centre-of-mass 
         frame,
         (c) $D^{\ast}$ pseudo-rapidity,
         (d) four-momentum transfer squared, $Q^2$. 
} 
\end{figure}

In Fig.~\ref{fig:difxsect}
the cross section results are presented as function of 
$p_{\perp}(D^{\ast})$, $p_{\perp}^*(D^{\ast})$, $\eta (D^{\ast})$, and $Q^2$.
Compared to the data are NLO QCD calculations 
performed with the HVQDIS program 
version 1.1~\cite{bh111} 
using a charm mass $m_c=1.5$ GeV.  
The parton density set CTEQ4F3~\cite{cteq} is used, which
is consistent with the Three Flavour Number Scheme
as implemented in HVQDIS. 
%and was shown in~\cite{bh111} to yield results 
%very close to those obtained with the GRV 94 HO set~\cite{grv-p}. 
%It is preferred for the subsequent analysis
%since 
The value for the QCD scale, 
$\Lambda^{(5)}=237$ MeV,
corresponds in the two-loop approximation the
world average for the strong coupling constant, 
$\alpha_s(M_{\rm Z}) = 0.119 \pm 0.002$~\cite{pdg}. 
For the $c\ra D^{\ast}$ fragmentation fraction, the value 27\%
is used, 
consistent with the average value of $(0.71\pm 0.05)\%$ for
the whole fragmentation and decay chain 
$c\ra D^{*+}\ra K^-\pi^+\pi^+$~\cite{ctoD}
and the value of 2.62\%~\cite{pdg} used for the decay branching fraction. 
For the fragmentation parameter
$\epsilon_c = 0.036$ is used,
which has been extracted~\cite{oleari}
from ARGUS $e^+e^-$ annihilation data using
NLO QCD in a Fixed Order massive charm scheme.

The NLO QCD predictions (displayed as histograms)
agree with the measurements except for $d\sigma / d\eta$
for which the calculations tend to undershoot the data at low $\eta$ and
to overshoot at high $\eta$. 
This disagreement can be accommodated within acceptable ranges of
parameters such as the QCD scale and parton density functions.
The shaded band in the figures represents the uncertainty 
of the theoretical prediction arising from different choices of the charm mass
between 1.3 and 1.7 GeV.

%================================================================

%gamma-p
\subsubsection{Photoproduction analysis}
Data are analyzed independently 
for the case 
where the scattered electron is detected in the 
electron tagger at 33 m and in the tagger
at 44 m.
Henceforth, the respective data samples will be referred to 
as {\it ETAG33} (recorded in 1994-96) and {\it ETAG44} (1995-96). 
The total integrated luminosities of these samples 
are $10.7~{\rm pb}^{-1}$ for {\it ETAG33} 
and $10.2~{\rm pb}^{-1}$ for {\it ETAG44}.
The two samples have no events in common. 
The kinematic ranges within which
the differential distributions are measured
are given in Tab.~\ref{tab:sigranges}.

The events were triggered by a coincidence of the {\it ETAG} 
signals with track candidate signals obtained from the MWPC
and central drift chamber trigger systems. 
Proton beam induced background is reduced by 
excluding events with energy flow only
into the forward region of the detector.

The acceptances of the electron taggers and their
trigger efficiencies are accounted for 
by assigning weights to individual events. 
The acceptance 
depends on $y$, 
which is reconstructed here using
$y = 1 - E_{e^{\prime}}/E_e$.
For the {\it ETAG33} sample it is above 20\% with
an average value of about 60\% in the specified range. 
This corresponds to a $\gamma p$ centre-of-mass energy range of
162\,\gev $ < W_{\gamma p} < 234$\,\gev,
with a  mean of $\langle W_{\gamma p}\rangle =194\,\gev$ and an average
\mbox{$\langle Q^{2}\rangle \approx 10^{-3}\,\gev^2$}. 
The acceptance of the {\it ETAG44} peaks sharply around $y=0.09$ 
and corresponds to
an average of $\langle W_{\gamma p}\rangle = 88~ \gev$.
For {\it ETAG44}, an average value for the acceptance is used for all events. 
 
The $K$ and $\pi$ candidates used in the $D^{\ast}$ reconstruction 
are required to have transverse momenta 
$p_{\perp} > 0.5\,\gev$ ({\it ETAG33}) or $p_{\perp} >
0.35\,\gev$ ({\it ETAG44}) respectively.
A cut of $p_{\perp} > 0.15\,\gev$ is imposed on the slow pion track. 
The method of 
{\it equivalent event numbers}~\cite{zech}
is used to determine the errors of 
the acceptance-weighted numbers of $D^{\ast}$ candidates.
The weighted numbers of combinations found in the $\Delta m$ fits are
$489 \pm 92$ (for {\it ETAG33}) and
$299 \pm 75$ (for {\it ETAG44}).

Using the efficiencies for the track component of the trigger condition, 
as well as the $D^{\ast}$ reconstruction and selection efficiencies 
as determined by Monte Carlo simulations, 
the cross sections 
for the specified ranges are found using Eq.~\ref{eq:xect}.
In the photoproduction case
QED radiative corrections are negligible due to the requirement of
no energy deposition in the photon detector of the luminosity system.
Systematic errors 
(see Tab.~\ref{tab:sigsyst}) 
are similar to those of the DIS analysis, 
except for those related to the electron measurement.
Here, uncertainties related to the electron tagger acceptance 
give a sizable contribution.  
The systematic errors are combined in quadrature to give $\pm 15\%$. 

Following
the {\em Weizs\"acker-Williams Approximation} (WWA)~\cite{wwa},
the electroproduction cross section $\sigma_{ep}$
is converted into a photoproduction cross section  
according to 
$\sigma(\gamma p \rightarrow D^{*}X) = \sigma (ep) / F$, 
where $F$ is the flux of photons emitted by the electron.
For {\it ETAG33} ({\it ETAG44}) the flux factor is 
$F=0.0128\, (0.0838)$. 

\begin{figure}[tbp]\centering
\unitlength1cm
\begin{picture}(17,14)
\put(0.0,7.5){\epsfig{file=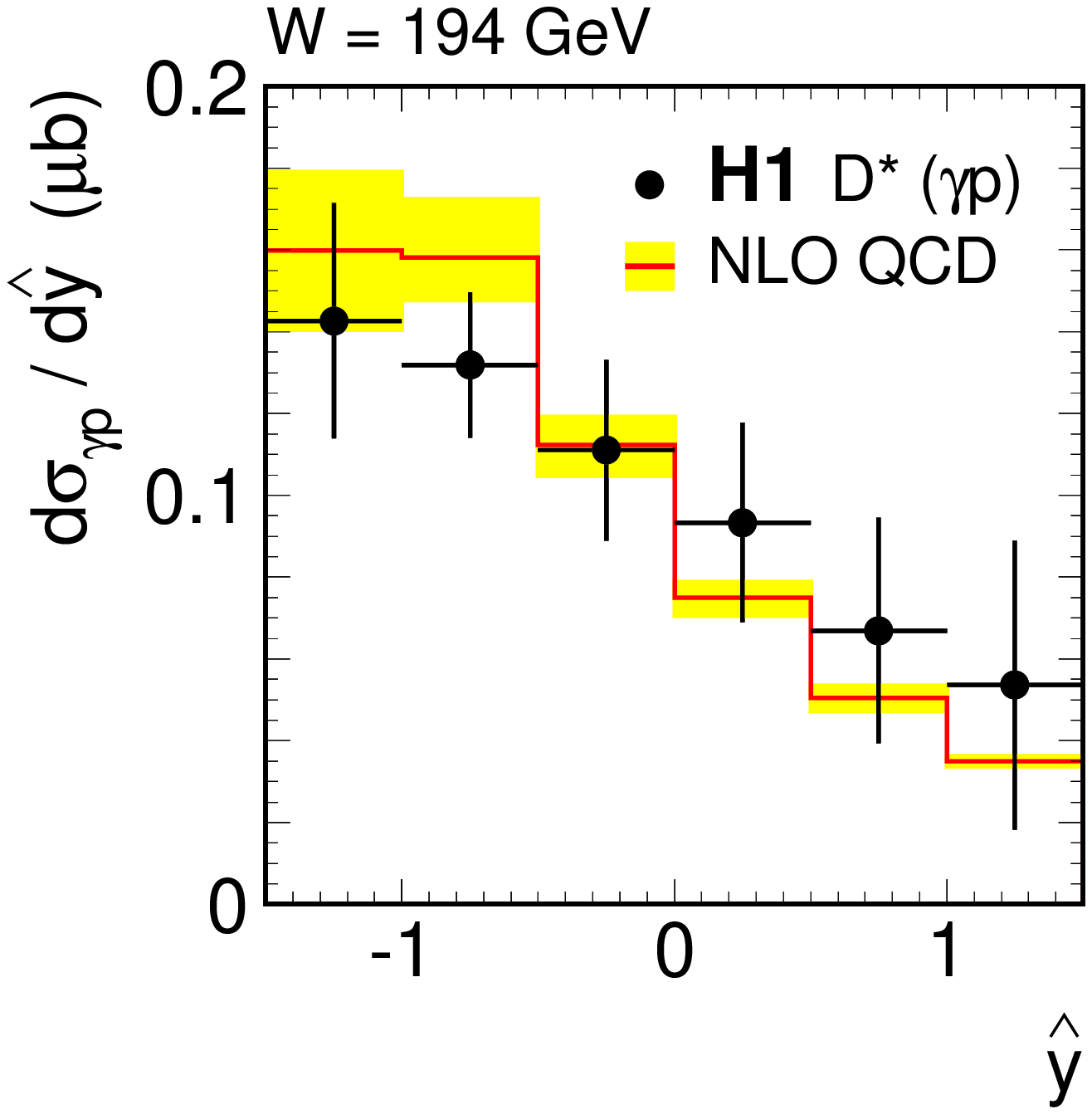,width=8cm}}
\put(8.,7.5){\epsfig{file=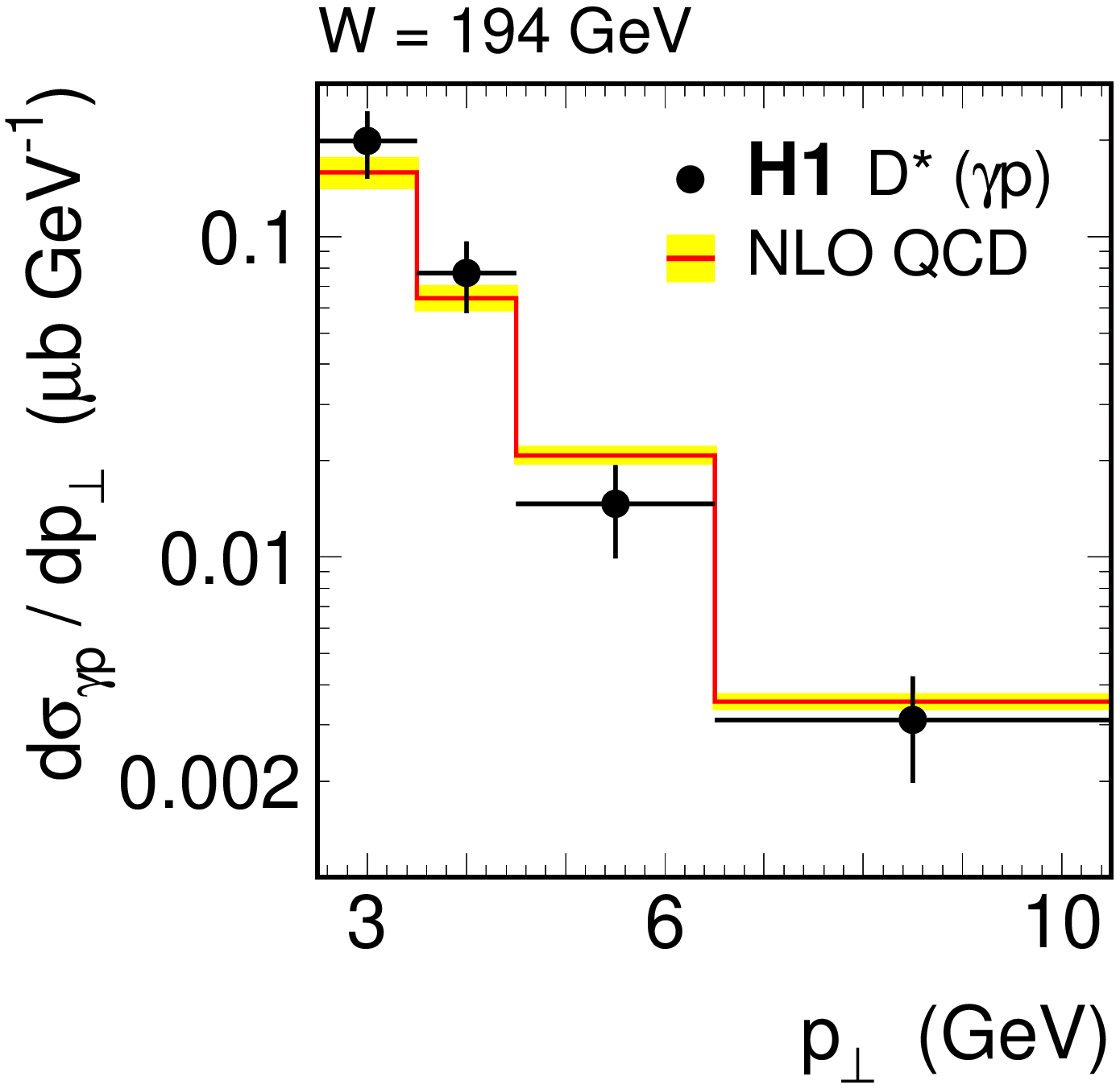,width=8cm}}
\put(0.0,0.0){\epsfig{file=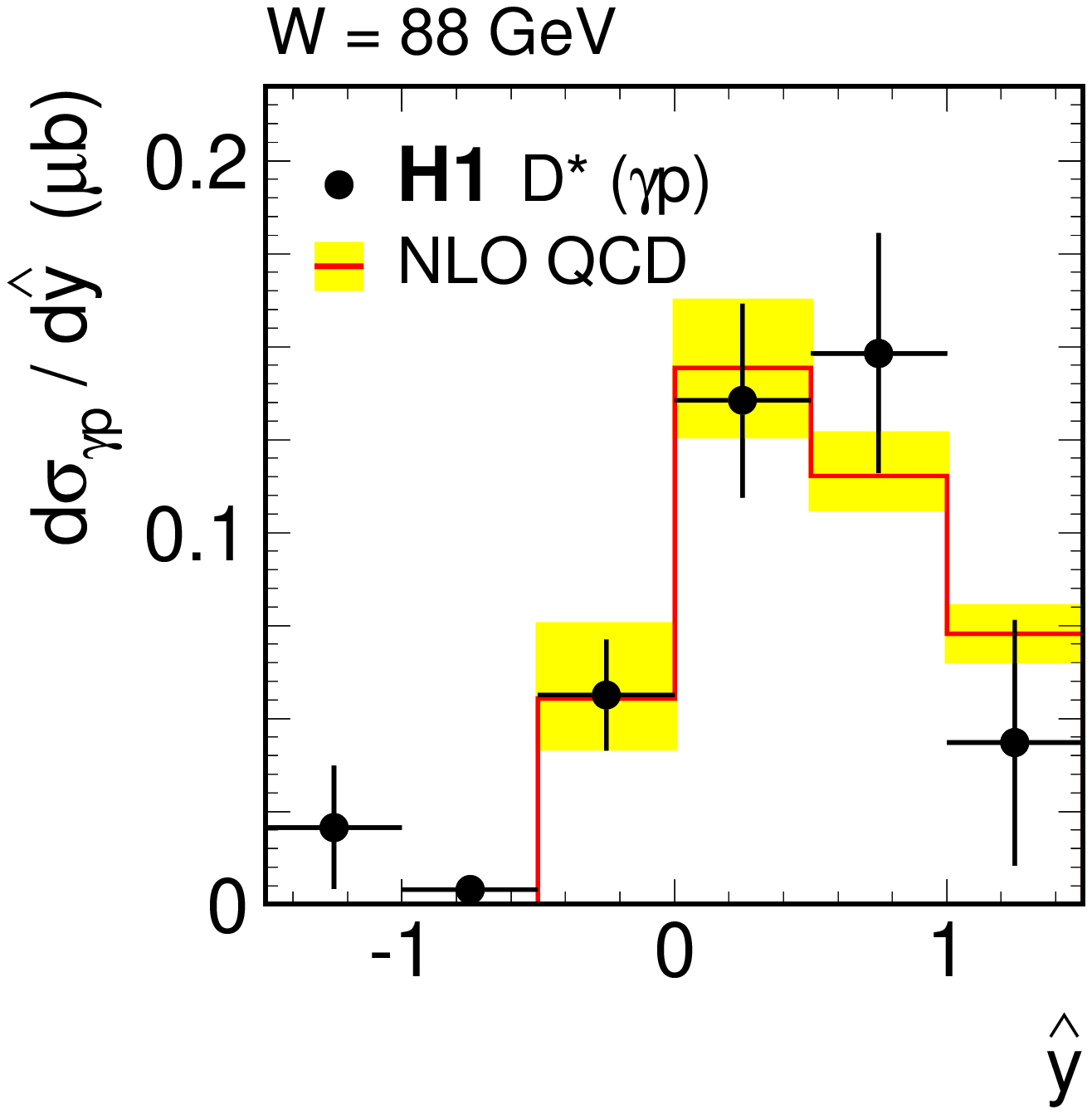,width=8cm}}
\put(8.,0.0){\epsfig{file=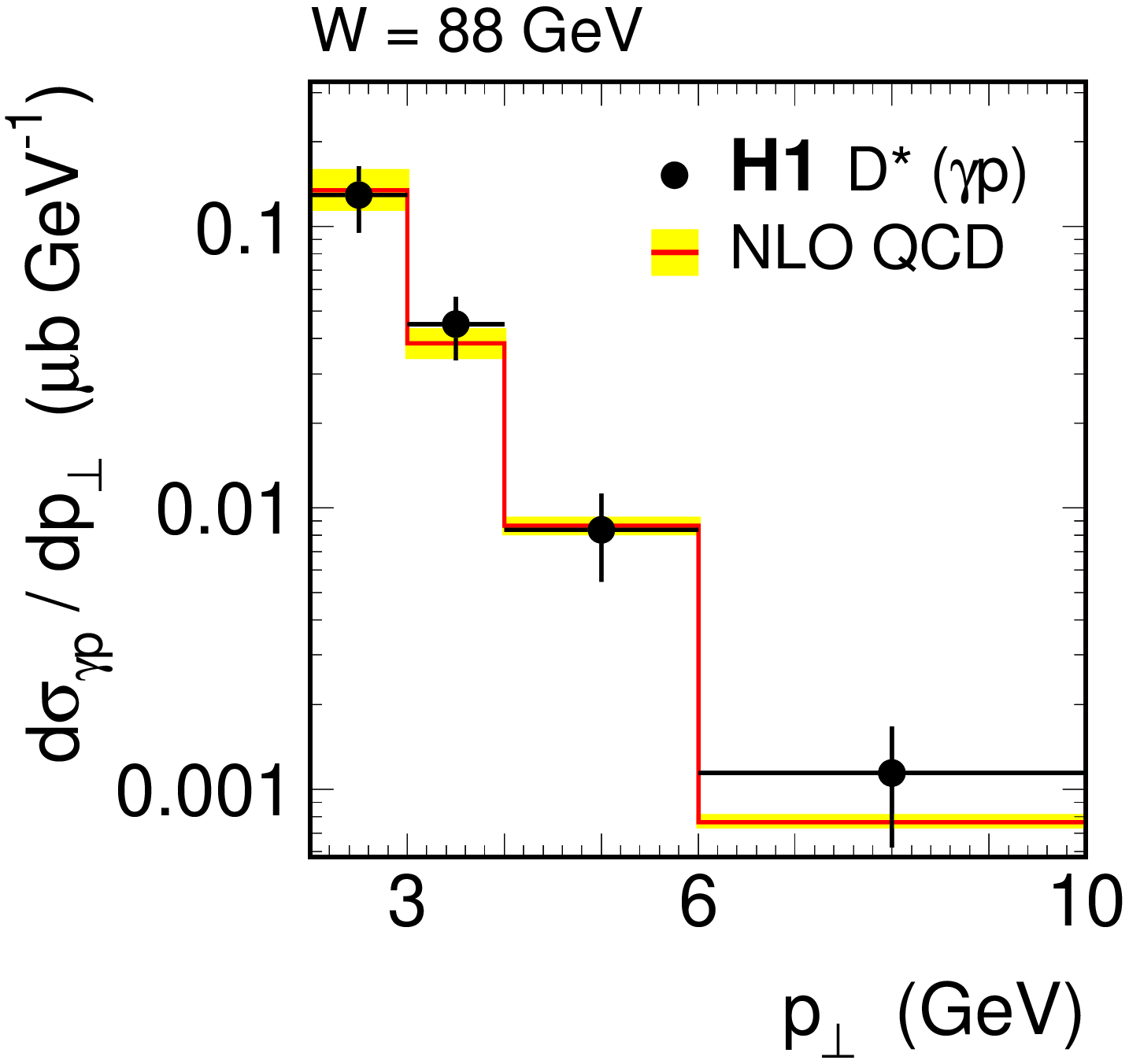,width=8cm}}
\put(2.3,9.5){a)}
\put(10.3,9.5){b)}
\put(2.3,3.){c)}
\put(10.3,2.){d)}
%\put(0,0){.(0,0)}
\end{picture}
\caption[Differential eta]%
        {\sl Differential photoproduction cross sections 
         in the kinematic range of experimental acceptance
         (see Tab.~\ref{tab:sigranges}).   
         The data are shown as solid dots with statistical error bars 
         for the ETAG33 sample
         %at an average $W_{\gamma p} = 194$ GeV 
         (a,b)
         and for the ETAG44 sample 
         %at $W_{\gamma p} = 88$ GeV 
         (c,d).
         A common systematic error of 15\% is not shown.
         The histograms show the NLO QCD predictions
         based on the FMNR program, using 
         the MRST1 
         parton density parameterization for the proton
         and the GRV-HO set for the photon. 
         The width of the shaded band 
         represents the variation of the theoretical cross sections 
         due to different choices of $m_c$ between 1.3 and 1.7 GeV.
  } 
\label{fig:si8384}
\end{figure}

\begin{figure}[p]\centering
\unitlength1cm
\begin{picture}(17,14)
\put(0.0,7.5){\epsfig{file=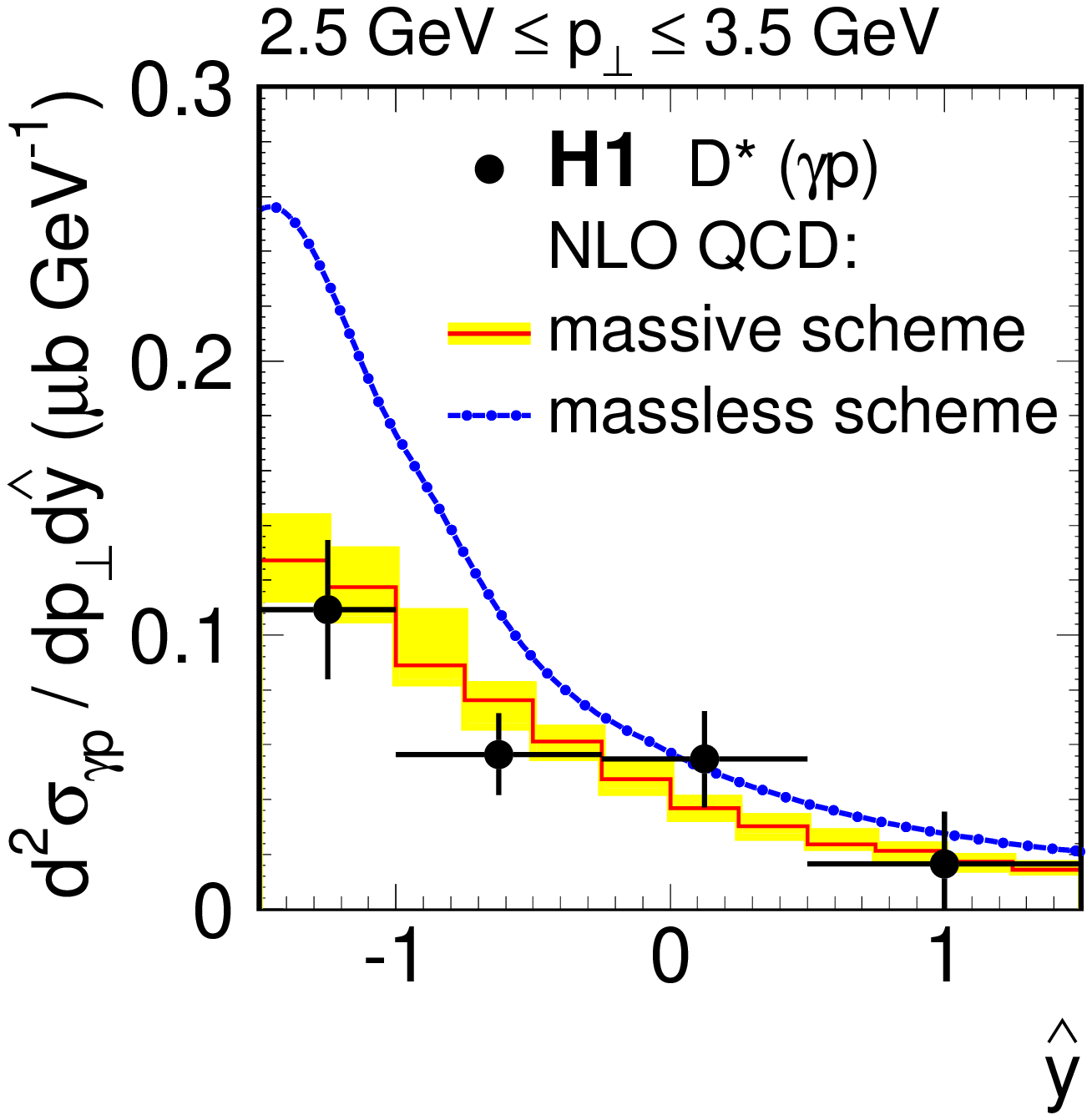,width=8cm}}
\put(8.,7.5){\epsfig{file=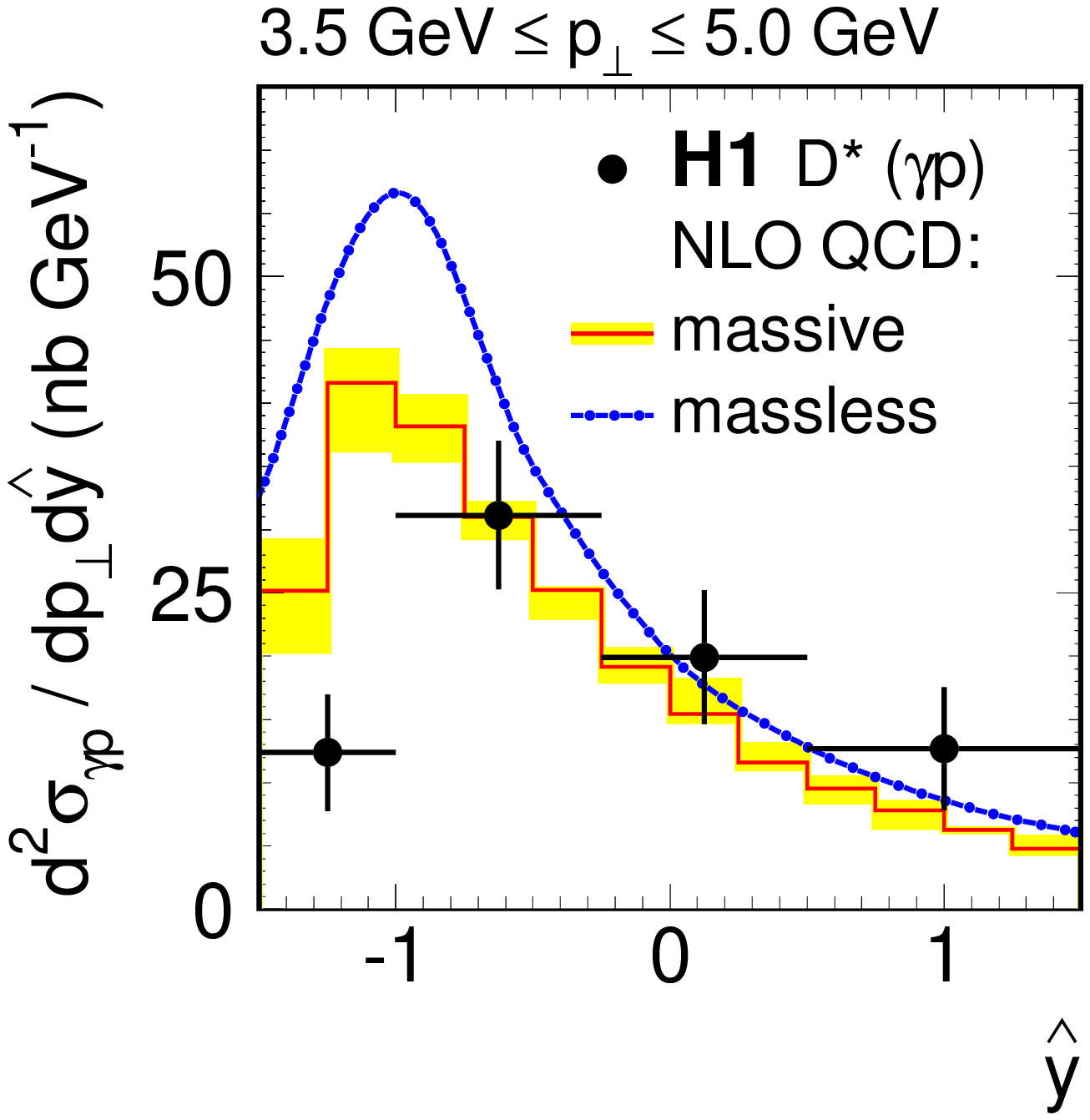,width=8cm}}
\put(4.,0.){\epsfig{file=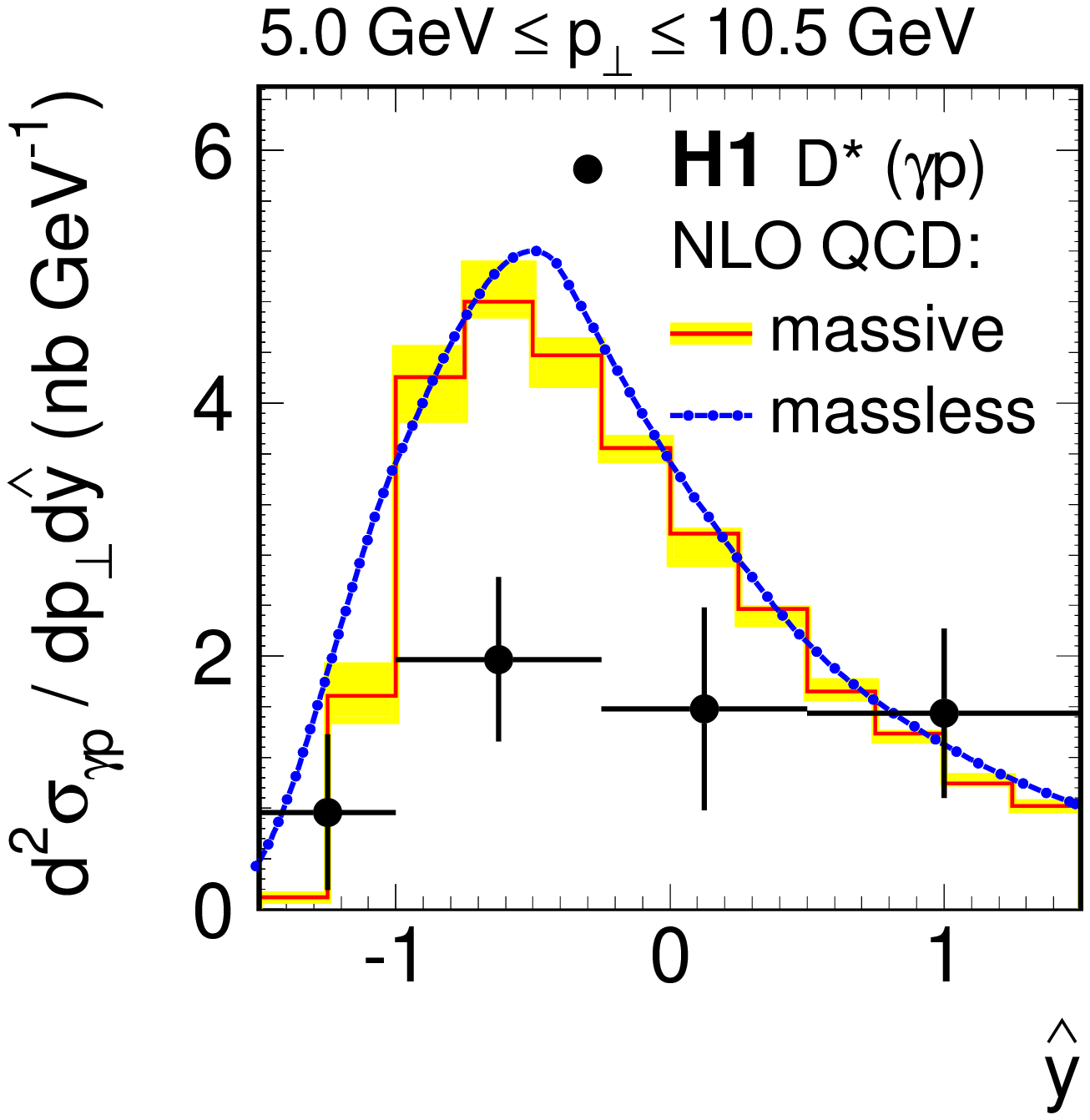,width=8cm}}
\put(2.3,9.3){a)}
\put(11.2,9.3){b)}
\put(7.3,1.8){c)}
%\put(0,0){.(0,0)}
%\put(0,0){.(0,0)}
\end{picture}
\caption[Differential eta]%
  {\sl Double-differential cross section 
  $d^2\sigma_{\gamma p} /d\hat{y} dp_{\perp}$   
  for three different ranges of transverse momentum
  at an average $W_{\gamma p} = 194$ GeV, shown as data points
  with statistical error bars.
  A common systematic error of 15\% is not shown.
  The histograms represent the NLO
  QCD predictions in the massive scheme based on the FMNR program 
  (using the parton density sets MRST1 (proton) and GRV-HO (photon)); 
  the shaded bands show their variations 
  due to different choices of the charm quark mass
  between 1.3 and 1.7 GeV.
  The curve shows the result of a NLO QCD calculation in the 
  massless scheme using the CTEQ4M parton density set for the proton
  and GRV-HO for the photon, 
  and a charm fragmentation function extracted from $e^+e^-$
  in the same framework.   
%In the low $p_{\perp}$ region,
%where the bulk of the events is found,
%the QCD calculation is expected to describe 
%most reliably the data.  
} 
\label{fig:sidd83}

\end{figure}

The measured distributions 
$d\sigma(\gamma p \rightarrow D^{*\pm}X) /d\hat{y}$ 
are shown 
for {\it ETAG33}
and for {\it ETAG44} 
in Fig.~\ref{fig:si8384}a and c, respectively,
for the range specified in Tab.~\ref{tab:sigranges}. 
The distributions  
$d\sigma_{\gamma p}/dp_{\perp} $ 
are shown 
in Fig.~\ref{fig:si8384}b,d. 
In addition, for {\it ETAG33} the double-differential distribution  
$d^2\sigma_{\gamma p}/d\hat{y} dp_{\perp}$  
is presented for three different ranges of transverse momentum in 
Fig.~\ref{fig:sidd83}.

The superimposed histograms shown in 
Fig.~\ref{fig:si8384} and in Fig.~\ref{fig:sidd83}
represent
the absolute predictions of the NLO QCD calculation\,\cite{frixione-1},
using, as in the DIS case,  
$m_c=1.5$ GeV, $\epsilon_c = 0.036$, and 27\% for 
the $c\ra D^{\ast}$ fragmentation fraction.  
The calculations are shown for the 
parton densities MRST1~\cite{mrst}
(proton) in combination with
GRV-HO~\cite{grv-g} (photon).
The histograms for the {\it ETAG33} sample are averages of calculations made
at three representative $W_{\gamma p}$ values, weighted
by the photon flux integrated over the represented range. 
For {\it ETAG44} the calculation is performed at 
a fixed $W_{\gamma p}=88$~GeV.
Reasonable agreement is observed for both
the shape and the absolute normalization between the 
measured single differential 
$p_{\perp}$  and $\hat{y}$ distributions and the NLO QCD calculation.
 
Fig.~\ref{fig:sidd83} reveals in more detail that the agreement is
good in the low $p_{\perp}$ region 
where the ``massive'' QCD calculation is expected to describe 
most reliably the data, 
and where the bulk of the events is found. 
For comparison, also overlaid in Fig.~\ref{fig:sidd83} 
is a calculation performed in the ``massless'' 
scheme~\cite{kniehlhera}. 
%for the region $p_{\perp}\gg m_c$, {\it i.e.\ } t
This approach is not expected to provide a description of 
the data in the region $p_{\perp}\sim m_c$.
Fig.~\ref{fig:sidd83} shows that the shape of 
the measured rapidity distribution 
in the upper $p_{\perp}$ range of this analysis
is less well described in both approaches.

\section{Determination of the Gluon Density}
\label{sec:glue}
\begin{figure}[p]\centering
\unitlength1.0cm
\begin{picture}(17,20)(0,0.4)
\put(0.0,0.){\epsfig{file=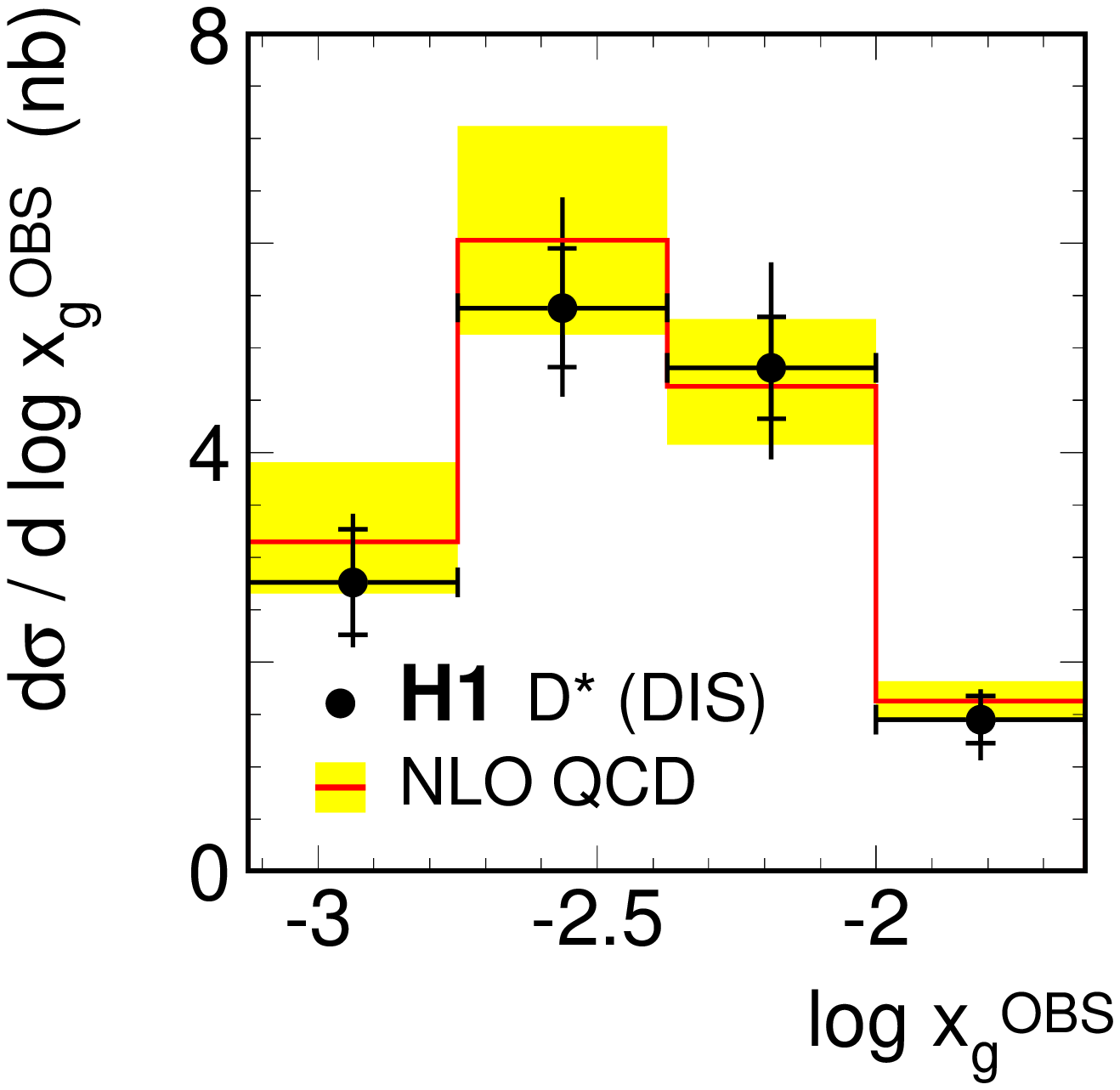,width=8cm}}
\put(8.,13.6){\epsfig{file=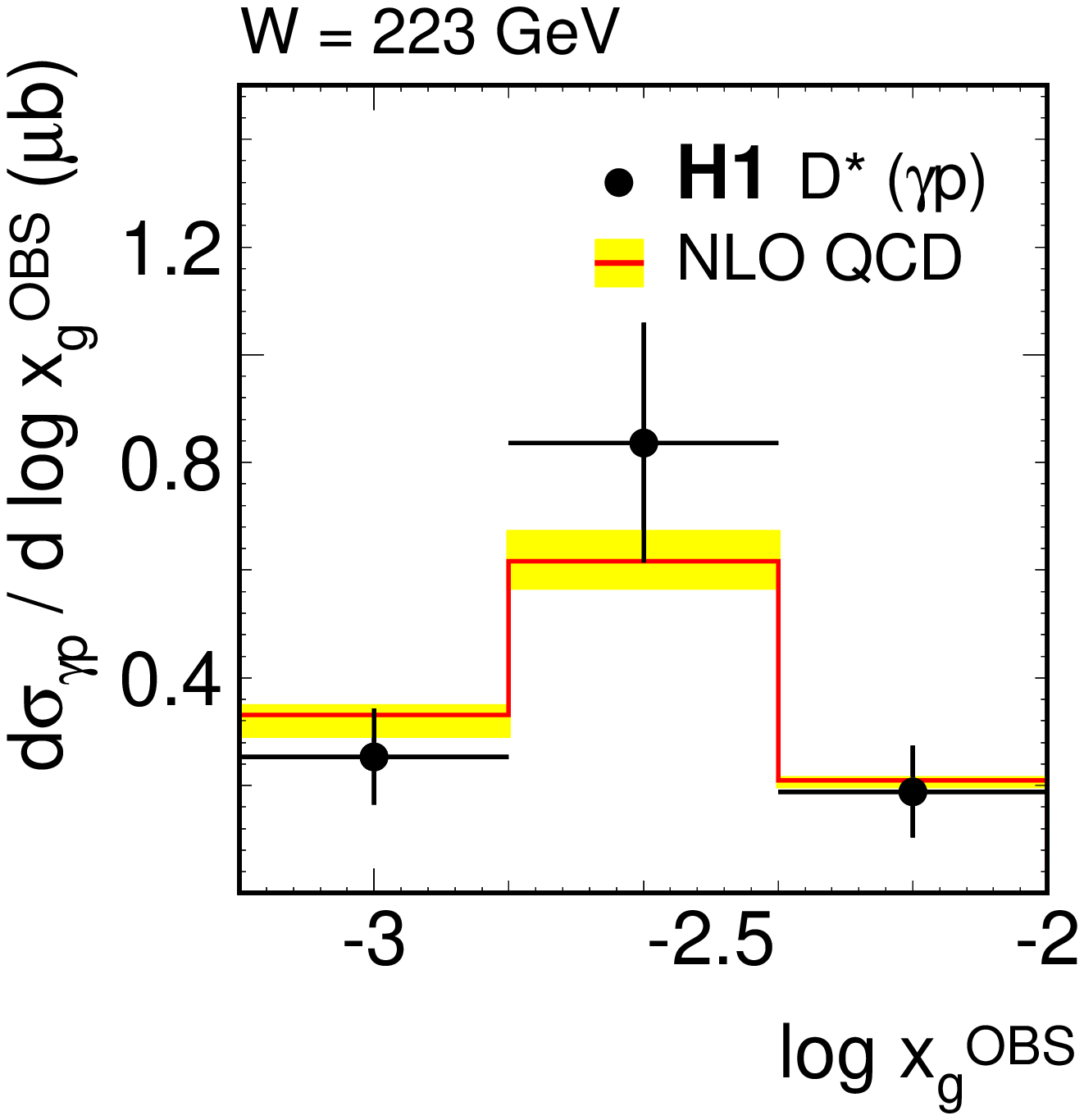,width=8cm}}
\put(8.,6.8){\epsfig{file=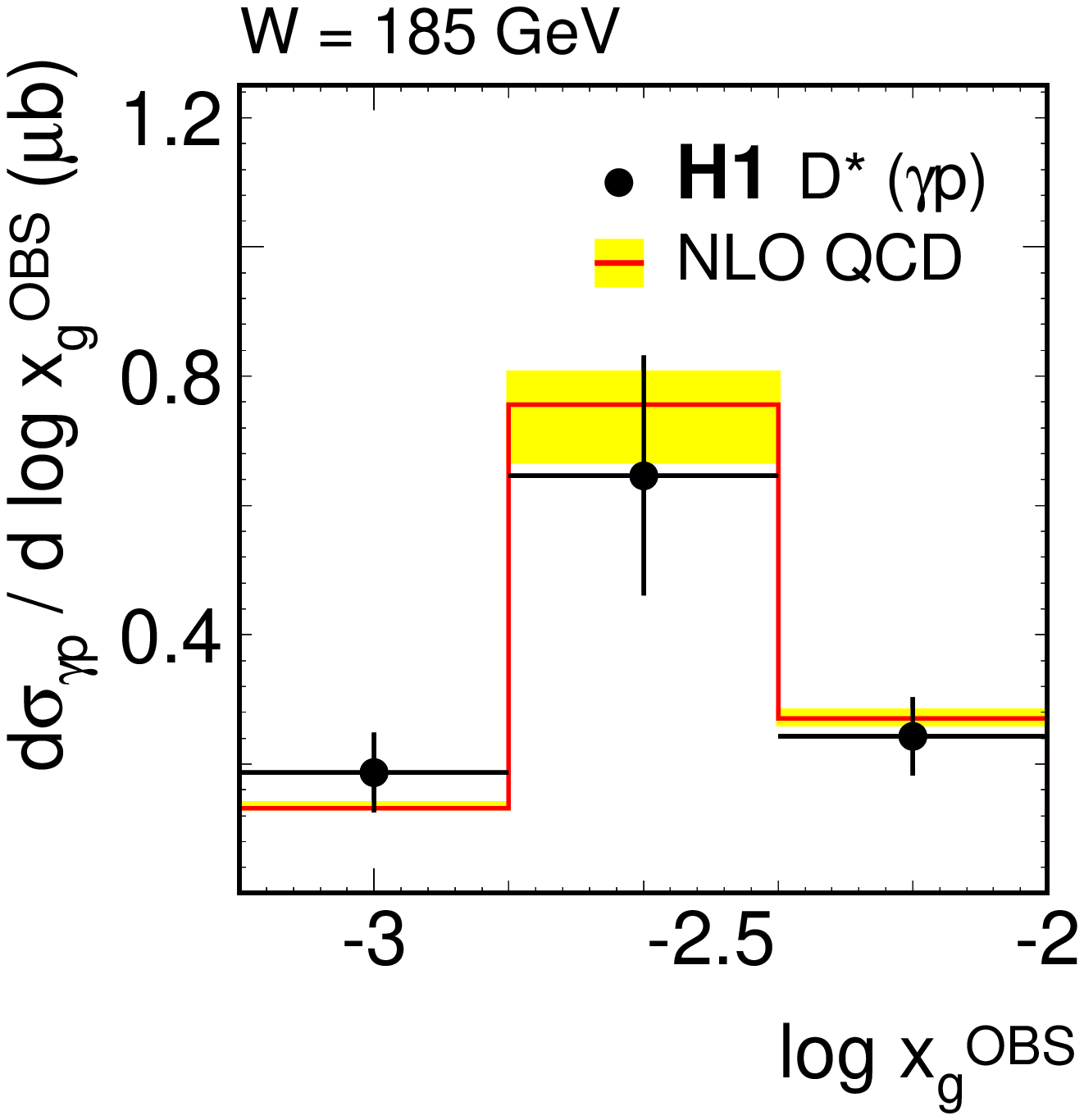,width=8cm}}
\put(8.,0.0){\epsfig{file=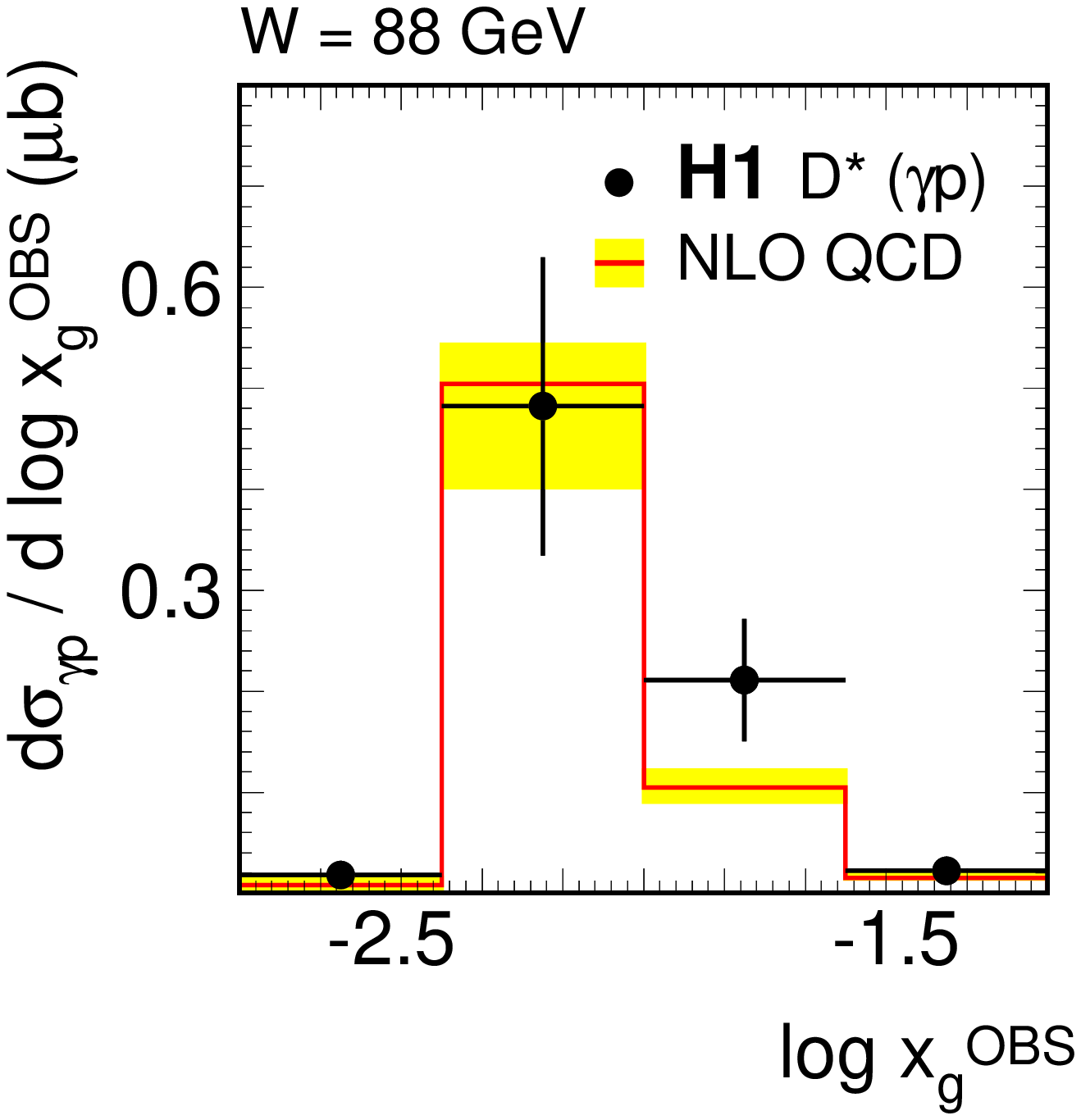,width=8cm}}
\put(2.3,5.7){a)}
\put(10.3,19.3){b)}
\put(10.3,12.5){c)}
\put(10.3,5.7){d)}
%\put(0,0){.(0,0)}
\end{picture}
\caption[dummy]{\label{fig:sigobs}\sl%
         (a) Differential DIS cross section          
         as a function of $x_g^{OBS}$ 
         in
         the kinematic range of experimental acceptance
         (see Tab.~\ref{tab:sigranges}).
         Figures (b-d) show the corresponding photoproduction cross sections, 
         separately for $W_{\gamma P}=223$, 185 and 88 GeV, respectively. 
         The meaning of the data points, histograms and their errors
         is the same as in Fig.~\ref{fig:difxsect} (for DIS) and 
         Fig.~\ref{fig:si8384} (for $\gamma p$).   
} 
\end{figure}

The gluon density $x_g g(x_g)$ is extracted from 
the measured cross sections $d\sigma /d(\log x_g^{OBS})$
which are compared to the NLO QCD predictions in Fig.~\ref{fig:sigobs}.

The photoproduction data (Fig.~\ref{fig:sigobs}b-d) are analyzed separately 
for three different photon-proton energies 
(88, 185, and 223 GeV).
Here, the rapidity range is restricted to  
$-1.5 \leq \hat{y} \leq 0.5$ (for {\it ETAG33})
and $0 \leq \hat{y} \leq 1$ (for {\it ETAG44}).
From the NLO calculation,
the resolved photon contribution
in this restricted range is expected 
to be less than 5\% 
(using the GRV-HO~\cite{grv-g} or LAC1~\cite{lac}
parton density set for the photon)
in every bin and is accounted for in the systematic error. 

% unfolding principle
The determinations of the gluon density in the proton 
from the DIS and from the photoproduction cross section
follow the same principles. 
The calculation of the differential $D^{\ast}$ cross section
in the experimentally accessible range
$\sigma (x_g^{OBS})$ 
can be written as
\begin{equation} \label{eqsigobs}
%\sigma (x_g^{OBS}) = \int dx_g \,\left [ g(x_g,\mu^2)
%                    \cdot \hat{\sigma} (x_g^{OBS},x_g,\mu^2) \right ]
\sigma (x_g^{OBS}) = \int dx_g \,\left [ g(x_g,\mu^2)
            \cdot \hat{\sigma} (x_g,\mu^2) A(x_g^{OBS},x_g,\mu^2) \right ]
                   + \sigma_{quark} (x_g^{OBS})
\end{equation}
where $\mu$ is the factorization scale, 
$\hat{\sigma}$ is the 
partonic cross section,
and $A$ is the integration kernel  
which contains the effects of gluon radiation
as well as fragmentation and
also incorporates 
the limited kinematic range of the measured cross section.
It receives the dominant contributions from 
regions where $x_g\approx x_g^{OBS}$.
The quark-induced contribution to the 
cross section (in the range of acceptance) 
is denoted $\sigma_{quark}$.
The form of Eq.~\ref{eqsigobs} is such that 
a determination of the gluon density in the proton
can be made from the measured cross section by means of 
an unfolding procedure.  

Firstly, according to Eq.~\ref{eqsigobs},
the quark-induced contribution
as predicted in the QCD calculation
is subtracted from the data bin by bin. 
It is about 1\% in total in DIS.
The relative contribution rises with $x_g^{OBS}$,
but remains below 10\% in the highest DIS  
and below 20\% in the highest $\gamma p$ bin.
The correlation with 
the gluon density via QCD evolution can be neglected here,  
but the uncertainty arising from it is included in the error.  

\begin{figure}[p]\centering
\unitlength1.0cm
\begin{picture}(17,14)
\put(-1.0,0.0){\epsfig{file=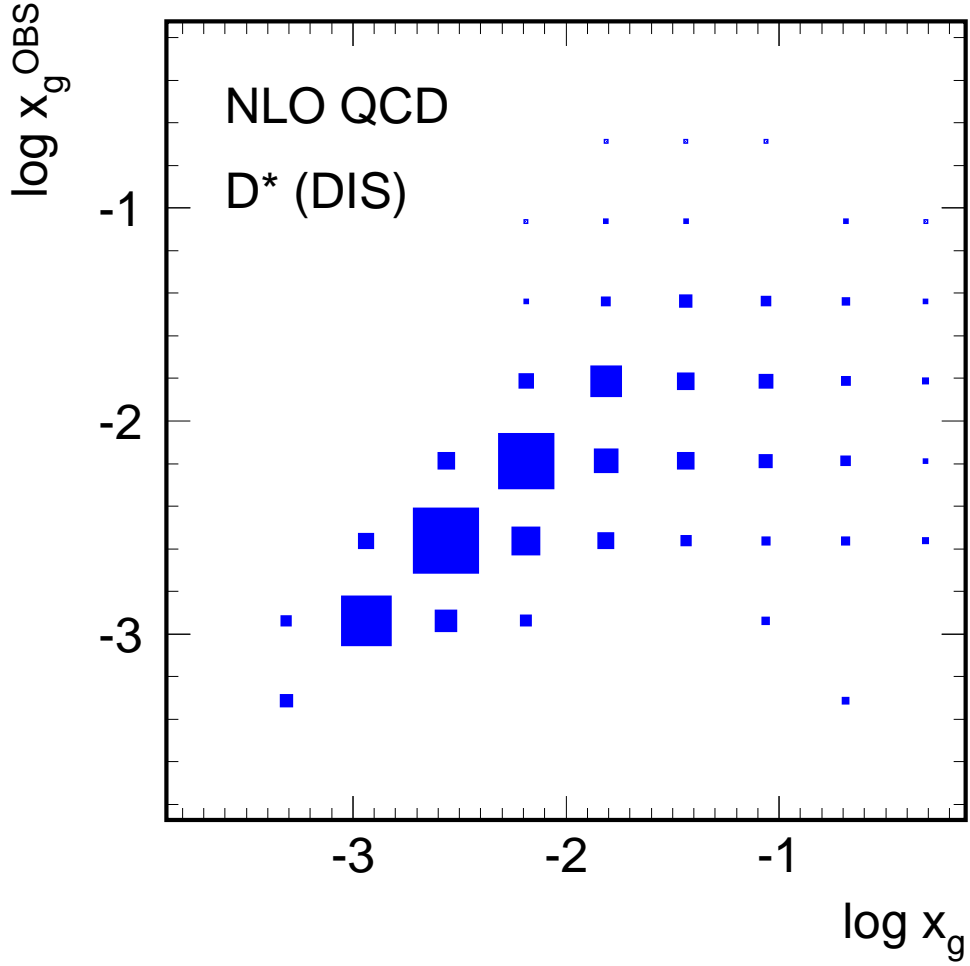,width=17cm}}
\end{picture}
\caption[dummy]{\label{fig:correl}\sl%
           Correlation between the observable $x_g^{OBS}$
           --- calculated using Eqs.~\ref{xgkine},\ref{repl} ---
           and the true momentum fraction $x_g$,
           as predicted in NLO QCD.
           The bin area corresponds to the cross section 
           contribution from the $x_g$ interval to the 
           observed $x_g^{OBS}$ range. 
           The correlation is distorted due to the effects of
           gluon radiation and fragmentation. 
} 
\end{figure}

The quark-subtracted cross sections 
are converted into cross section distributions as a function of the true 
$x_g$ using the iterative unfolding procedure for binned 
distributions of Ref.~\cite{dagostini}. 
The effects of limited detector resolution on $x_g^{OBS}$
are small with respect to the bin size and have already 
been corrected for in the measured cross section. 
The correlation between $x_g^{OBS}$ 
and $x_g$ 
as obtained from the QCD calculation is shown for the DIS case 
as a two-dimensional histogram in Fig.~\ref{fig:correl}.
This correlation is used for the first unfolding iteration
and re-weighted for each subsequent iteration   
using the result of the previous one. 
After 4 steps at most, convergence is reached, 
and the ``smeared" distribution $\sigma (x_g^{OBS})$ 
agrees with the measurement.

\begin{figure}[p] \centering
\unitlength1.0cm
\begin{picture}(17,14)
\put(-1.0,0.0){\epsfig{file=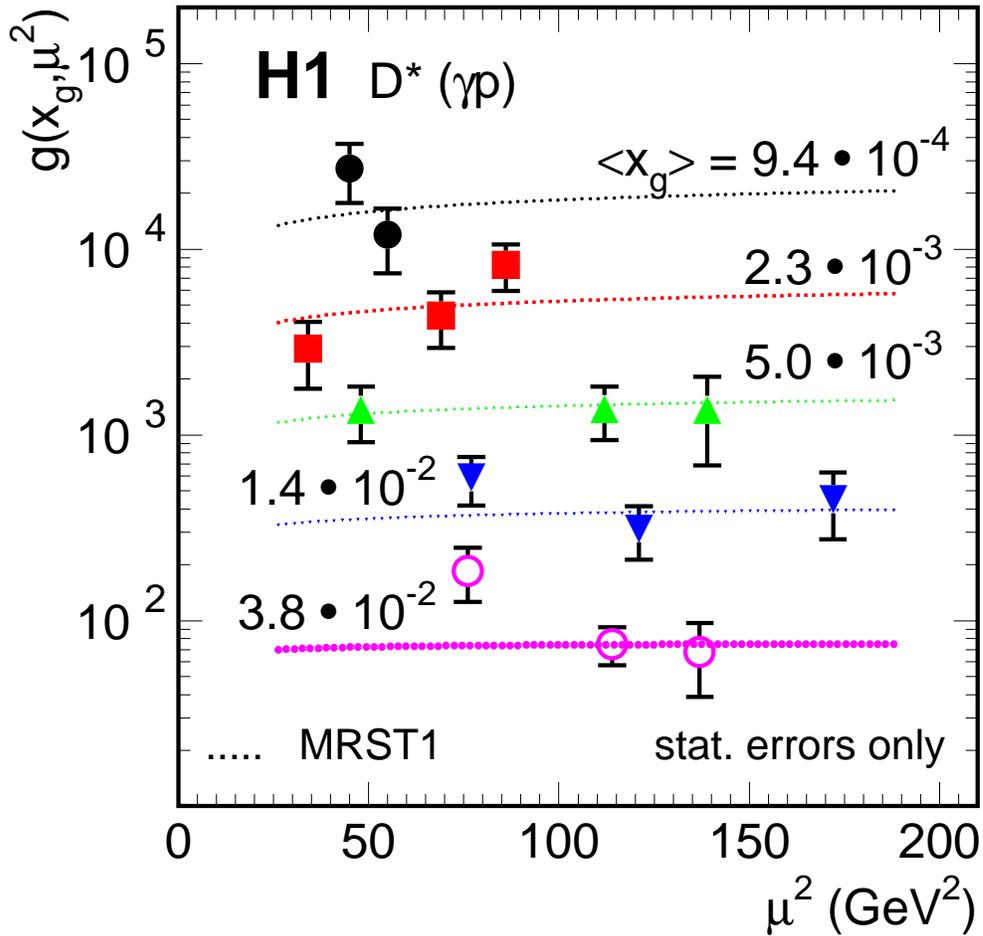,width=17.cm}}
\end{picture}
\caption[m]
        {\sl Gluon density distribution, 
         measured in photoproduction 
         as a function of the scale $\mu^2$.
         Data are plotted as $g(x_g,\mu^2)$ and not $x_gg(x_g,\mu^2)$,
         in order to show most clearly the evolution with scale.  
         Each key symbol represents a different
         bin in $x_g$ with the average value given in the figure. 
         The error bars represent the statistical errors only.
         The dotted
         lines represent the gluon density of the MRST1 parameterization,
         shown as a function of scale $\mu^2$ for the same set of $x_g$ values. 
}
\label{fig:gmu}
\end{figure}

The unfolded cross section distribution factorizes: 
\begin{equation}  
\sigma (x_{g,i}) \sim
g(x_{g,i},\mu _i^2) \cdot  \hat{\sigma}(x_{g,i},\mu _i^2) \; ;
\end{equation}
but for every bin $i$ in $x_g$ (and $W$) the
gluon density is probed at a different factorization scale $\mu_i$
which depends on the phase space region.
%$\hat{\sigma}$ is a partonic cross section. 
The scale in the DIS measurement changes from 
17\, GeV$^2$ in the first to 28\, GeV$^2$ in the last bin
in $x_g$, 
the average being 
$\langle\mu^2\rangle = 25$ GeV$^2$.
The $Q^2$ dependence of the cross section is well reproduced
(see Fig.~\ref{fig:difxsect}d).
In photoproduction the factorization scale varies between 
30 and 140 GeV$^2$, with an average value of 50 GeV$^2$,
as can be seen in Fig.~\ref{fig:gmu} where $g(x_g)$
as determined from the photoproduction data is 
plotted as a function of the scale $\mu$
associated to the $x_g$ and $W$ range of each measurement.
To determine $g(x,\langle\mu^2\rangle )$ at an average scale,
all data points are scaled to an average $\langle \mu^2\rangle$: 
\begin{equation} \label{eq:gsig}
g(x_{g,i},\langle\mu^2\rangle ) =  
                \frac{g(x_{g,i}, \langle\mu^2\rangle )^{\mbox{\small theory}}}
                     {g(x_{g,i},\mu _i^2)^{\mbox{\small theory}}}
   \,\cdot\,    \frac{g(x_{g,i}, \mu^2_i)^{\mbox{\small theory}}}
                     {\sigma (x_{g,i})^{\mbox{\small theory}}}
            \,\cdot\,  \sigma (x_{g,i})^{\mbox{\small experiment}} \;\; .
\end{equation}
The change of $g(x_{g,i},\mu^2)$ 
when varying the factorization scale from $\mu_i^2$ to $\langle\mu^2\rangle$
is taken from theory, assuming the same scale dependence as in CTEQ4
(DIS) or MRST1 ($\gamma p$). 
This change is however small: in the DIS measurement
$g(x, \mu^2_i)$  deviates from $g(x, \langle\mu^2\rangle )$ 
by no more than $9\%$. 
In photoproduction, the variation can be inferred
from Fig.~\ref{fig:gmu}. 
The effect is included in Eq.~\ref{eq:gsig}, 
and the dependence on the input parton density scale variation is 
accounted for in the systematic error. 

\begin{figure}[p]\centering
\unitlength1.0cm
\begin{picture}(17,19)(0,1.1)
\put(1.,9.7){\epsfig{file=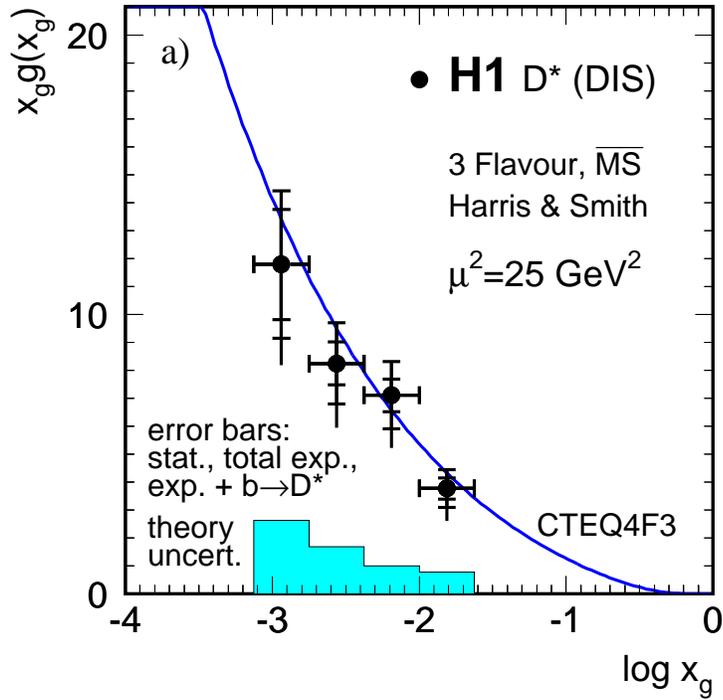,width=12.5cm}}
\put(1.,0.){\epsfig{file=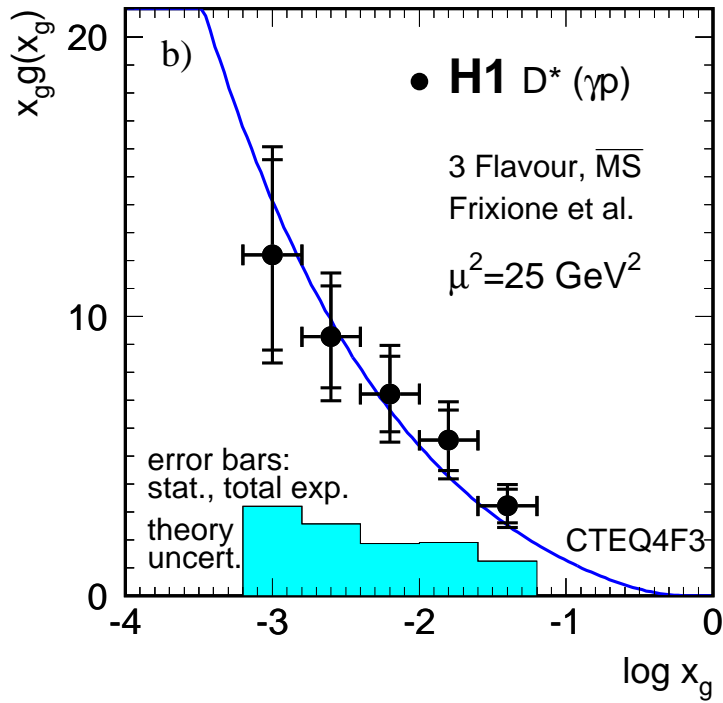,width=12.5cm}}
\put(4.6,19.1){\large a)}
\put(4.6,9.4){\large b)}
%\put(0,0){.(0,0)}
\end{picture}
\caption[dummy]{\label{fig:gluon}\sl%
         Gluon density, determined from DIS (a)
         and from from photoproduction data (b),
         compared to the CTEQ4F3 parameterization at $\mu^2=25$
         GeV$^2$ (full line). 
         The error bars re\-pre\-sent the statistical (inner) and 
         total experimental (outer) error; 
         in addition the effect of a subtraction 
         of $b\bar{b}$ background (if 5 times higher than predicted in 
         a LO Monte Carlo program) 
         is displayed in (a),
         in (b) the $b\bar{b}$ contribution
         is negligible.
         The theoretical systematic error
         is shown as shaded band along the abscissa
         and is dominated in DIS (a) by the uncertainty associated 
         with the charm quark mass, 
         in photoproduction (b) 
         it is dominated by the uncertainty  
         associated 
         with the renormalization and factorization scale.
}
\end{figure}

The resulting gluon distribution, $x_g g(x_g)$, 
is shown in Fig.~\ref{fig:gluon}a
for the DIS case at the average scale of $\mu^2=25$ GeV$^2$. 
The covariance matrix for the statistical uncertainty
has been calculated;
the inner error bars in the figure represent the diagonal elements only.
The correlation coefficients
(defined as 
$\rho _{ij} = \mbox{cov}_{ij}/\sqrt{\mbox{cov}_{ii}\mbox{cov}_{jj}})$
are 18\%, 27\%, 42\% for the first, second and third pair of 
neighbouring bins, respectively.
An overall normalization uncertainty arises because of the experimental 
systematic error of the $D^{\ast}$ cross section. It is increased
by the additional uncertainty from the $c\ra D^{\ast}$ fragmentation fraction,
and is added (in quadrature) to the statistical error 
to give the total experimental error shown as 
outer error bars. 
The downward extension of the error bar in Fig.~\ref{fig:gluon}a
represents the decrease of $x_gg(x_g)$
if the $b\bar{b}$ contribution to the $D^{\ast}$ 
cross section is 5 times the size predicted by the AROMA Monte Carlo.
The corresponding result from the photoproduction analysis 
is shown in Fig.~\ref{fig:gluon}b.

Theoretical uncertainties affecting the gluon density extraction 
are determined by re-cal\-cu\-la\-ting with the QCD programs for different
parameter choices the cross sections and the correlations needed 
in the unfolding procedure. 

The dominant contribution in the DIS case arises from the uncertainty
due to different choices of the charm quark mass between 1.3 and 1.7 GeV.  
In addition, 
the QCD factorization and renormalization scale $\mu$
was varied between $2m_c$ and $2\sqrt{4m_c^2+Q^2}$ in DIS,
following~\cite{bh111}.

For the photoproduction results,  
the scale variation dominates the theoretical uncertainty 
over the whole range; 
a factor 2 is allowed in both directions.
Here the charm quark mass uncertainty affects mainly the scale
and is not added as a separate contribution. 

One of the larger uncertainties 
is due to our ignorance of the charm fragmentation function 
which has not yet been measured in $ep$
scattering. 
It is assumed here that this can be taken from measurements in 
$e^+e^-$ annihilation. 
A fragmentation parameter $\epsilon_c$ range between 0.026 and 0.046
is allowed in the calculation. 
Included in this range are the values extracted from 
ARGUS and OPAL data in a Fixed Flavour Number scheme~\cite{oleari}. 
The average uncertainty in $x_gg(x_g)$ 
due to this variation of the fragmentation parameter 
is 4\% in DIS and 10\% 
in photoproduction.

The uncertainty in the strong coupling constant (world average) 
is taken into account by variation of the scale $\Lambda^{(5)}$ between
213 and 263 MeV~\cite{pdg}.
In order to assess the dependence on the
input parton density parameterization, 
the MRS(A')~\cite{mrsa} set is used.
In photoproduction, 
an additional error of $^{-4\%}_{+0\%}$ 
accounts for resolved contributions and the uncertainty
in the photon structure function% 
\footnote{In DIS, resolved contributions 
are expected to be negligible~\cite{grs}.}.
 
The contributions to the theoretical error due to the discussed
variations in the parameters of the NLO QCD calculations are 
added in quadrature and displayed 
as shaded histograms along the abscissa in 
Fig.~\ref{fig:gluon}.
The errors at small $x_g$ are dominated by the statistical uncertainty.

\begin{figure}[p]\centering
\unitlength1.0cm
\begin{picture}(17,14)
\put(-1.0,0.0){\epsfig{file=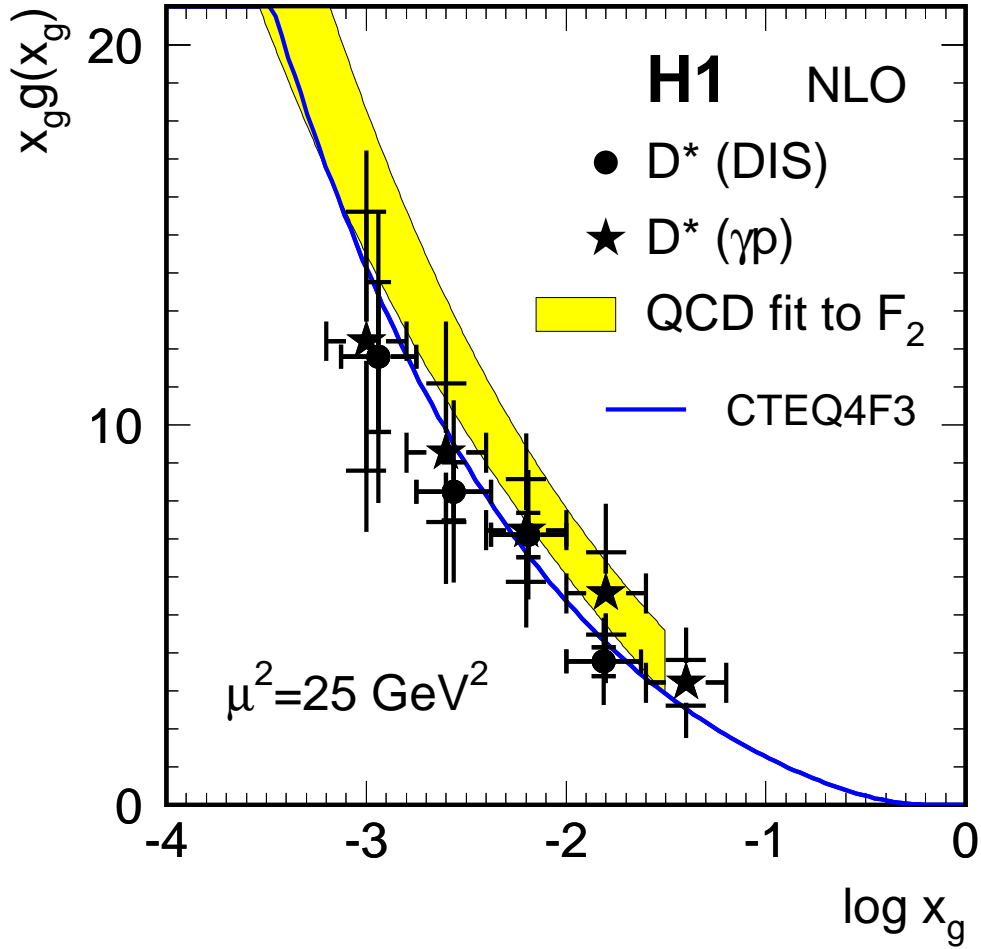,width=17cm}}
\end{picture}
\caption[dummy]{\label{fig:both}\sl%
  Gluon densities obtained from the two
  $D^{\ast}$ analyses. 
  The inner error bars represent the statistical and the outer
  the total error.
  The systematic error is a
  quadratic sum of all contributions, dominated by the theoretical 
  uncertainty.
         Both results are compared 
         to the result of the H1 QCD analysis
         of the inclusive $F_2$ measurement~\cite{h1f2}
         at $\mu^2=25$ GeV$^2$
         (light shaded band) and the CTEQ4F3 parameterization.                 
}
\end{figure}

The gluon densities 
extracted from DIS and photoproduction cross sections
are in very good agreement with each other,
as can be seen in Fig.~\ref{fig:both}.
% where they are both shown at the same scale $\mu^2=25$ GeV$^2$. 
They can be compared with other results determined 
in NLO from other processes. For example, 
the result for $x_g g(x_g)$ 
determined by H1~\cite{h1f2}
from a QCD analysis of the scaling violations of $F_2$ 
is also overlaid in Fig.~\ref{fig:both}.
This result has been obtained from a fit
to an inclusive measurement of the proton structure function $F_2$
using the NLO QCD evolution equations
and the boson-gluon fusion prescription
for the charm treatment. 
Within errors 
it agrees 
well with the gluon distributions
determined in the analyses presented here, which are
performed in the framework of the Three Flavour Number Scheme for DIS
as implemented in the HVQDIS program,
and in the ``massive charm'' calculations 
for photoproduction,
implemented in the FMNR program.
This amounts to an important verification of our understanding of the
application of QCD in low$-x$ DIS and to a demonstration of the
universality of the gluon distribution in the proton within the framework
of such calculations at NLO.

\section{Conclusion}

Differential 
$D^{\ast}$ cross sections 
have been measured in deep inelastic
$ep$ scattering 
and in photoproduction.  
The results have been presented as a function of
various kinematic variables and 
in different kinematic ranges.
They have been compared to NLO QCD calculations 
in the ``massive'' Three Flavour Number Scheme.
The calculations were performed  
using the programs HVQDIS and FMNR, 
which have implemented 
the Peterson form of the fragmentation function
as determined from $e^+e^-$ annihilation data
to model the non-perturbative hadronization process.
The overall description of the differential cross sections by 
the theoretical calculations has been found to be adequate.

Using the explicit NLO QCD calculations, 
the gluon density in the proton, $x_g g(x_g)$
has been determined 
from the measured cross sections as a function of $x_g^{OBS}$.
The experimental observable $x_g^{OBS}$ 
is used as an estimator for the gluon momentum fraction $x_g$ 
using the kinematic information from
the $D^{\ast}$ meson in addition to the scattered electron.
The function $x_g g(x_g)$ has been extracted 
at an average scale $\mu ^2=25$ GeV$^2$.
The gluon densities obtained from 
electroproduction and from photoproduction of charm 
agree well with each other and 
with that derived from
the QCD analysis of the inclusive measurement 
of the proton structure function $F_2$.

\section*{Acknowledgments}

We are grateful to the HERA machine group whose outstanding
efforts have made and continue to make this experiment possible. 
We thank
the engineers and technicians for their work in constructing and now
maintaining the H1 detector, our funding agencies for 
financial support, the
DESY technical staff for continual assistance, 
and the DESY directorate for the
hospitality which they extend to the non-DESY 
members of the collaboration.
We wish to thank S.~Frixione, B.~Kniehl, 
B.~Harris and J.~Smith for 
support in the QCD calculations, 
and P.~Nason and C.~Oleari for advice on fragmentation issues. 

%Figures at the end of the paper if possible please

% the bibliography

%\section*{Figures}
\newpage


\end{document}